\documentclass[vecphys]{svmult}

\usepackage{times}
\usepackage{url}
\usepackage{color}
\usepackage{makeidx}         
\usepackage{graphicx}        
\usepackage{multicol}        
\usepackage[bottom]{footmisc}

\makeindex             

\begin{document}

\setcounter{chapter}{0}

\title{Reaction kinetics of coarse-grained equilibrium polymers: a Brownian Dynamics study}
\titlerunning{Equilibrium polymers}
\author{Chien-Cheng Huang \inst{1,2} 
\and Hong Xu \inst{2} \and Fran\c{c}ois Crevel \inst{3} 
\and Joachim Wittmer \inst{3} \and Jean-Paul Ryckaert \inst{1}}
\authorrunning{C-C Huang \and H. Xu \and  F. Crevel \and J. Wittmer \and J.-P. Ryckaert} 
\institute{Physique des polym\`eres, CP223, Universit\'{e} Libre de Bruxelles,
Bv du Triomphe, 1050 Brussels, Belgium
\texttt{jryckaer@ulb.ac.be}
\and  LPMD, Inst. Physique-Electronique, Univ. Paul Verlaine-Metz,
1bd Arago, 57078 Metz cedex 3, France
\texttt{hongxu@univ-metz.fr}
\and Institut Charles Sadron, 6 Rue Boussingault,67083 Strasbourg, France
\texttt{jwittmer@ics.u-strasbg.fr}
}
%
%
\maketitle

\section{Abstract}
Self-assembled linear structures like giant cylindrical micelles or discotic
molecules in solution stacked in flexible columns are systems reminiscent of
polydisperse polymer solutions, ranging from dilute to concentrated solutions
as the overall monomer density and/or the chain length increases. These supramolecular polymers
have an equilibrium length distribution, the result of a competition between 
the random breakage of chains and the fusion of chains to generate longer ones.
This scission-recombination mechanism is believed to be responsible of some 
peculiar dynamical properties like the Maxwell fluid rheological character for entangled 
micelles. Simulations employing simple mesoscopic models provide a powerful
approach to test mean-field theories or scaling approaches which have been
proposed to rationalize the structural and kinetic properties of these soft
matter systems. In the present work, we review the basic theoretical concepts 
of these ``equilibrium polymers" and some of the important results obtained by 
simulation approaches. We propose a new version of a mesoscopic model in continuous 
space based on the bead and FENE spring polymer model which is treated by Brownian 
Dynamics and Monte-Carlo binding/unbinding reversible changes for adjacent
monomers in space, characterized by an attempt frequency parameter $\omega$.
For a dilute and a moderately semi-dilute state-points which both correspond to
dynamically unentangled regimes, the dynamic properties are found to
depend upon $\omega$ through the effective life time $\tau_b$ of the average size
chain which, in turn, yields the kinetic reaction coefficients of the mean-field 
kinetic model proposed by Cates. Simple kinetic theories seem to work for times $
t \geq \tau_b$  while at shorter time, strong dynamical correlation effects are observed.
Other dynamical properties like the overall monomer diffusion and the mobility of reactive 
end-monomers are also investigated.


\section{Introduction}

\begin{figure}[th] 
\centerline{\includegraphics[width=.5\textwidth]{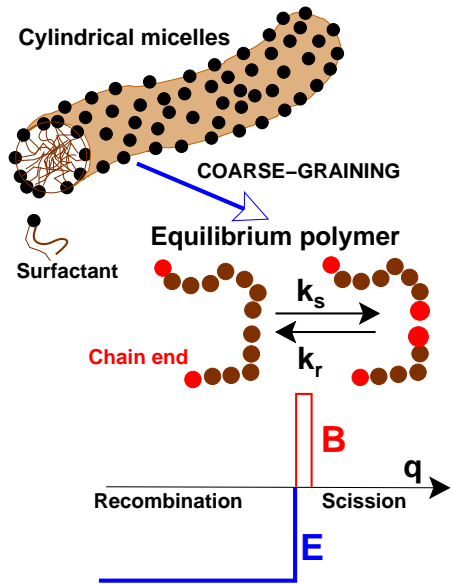}}
\vspace*{12pt} \caption{Some surfactant molecules in solution self-assemble and form
long wormlike micelles which continously break and recombine.Their mass distribution 
is, hence, in thermal equilibrium and they present an important example of the vaste 
class of systems termed ``equilibrium polymers"\protect\cite{VdS}.The free energy $E$ of the 
(spherical) end cap of these micelles has been estimated \protect\cite{catc} to be of order of 
$10k_BT$. This energy penalty (together with the monomer density) determines essentially 
the static properties and fixes the ratio of the scission and recombination rates, 
$k_s$ and $k_r$. Additionally, these rates are influenced by the 
barrier height $B$ which has been estimated to be similar to the end cap energy. 
Both important energy scales have been sketched schematically as a function of a generic 
reaction coordinate $q$ (see chapter 8 of reference \protect\cite{chand}).Following closely 
the analytical description \protect\cite{catc,witt98} these micellar systems are 
represented in this study by coarse-grained effective potentials in terms of a standard 
bead-spring model. The end cap free energy becomes now an energy penalty for scission 
events, i.e., the creation of two unsaturated chain ends. The dynamical barrier is 
taken into account by means of an attempt frequency $\omega = \exp{(-B/k_BT)}$. If 
$\omega$ is large, successive breakage and recombination events for a given chain can 
be assumed to be uncorrelated and the recombination of a newly created chain ends will 
be of standard mean-field type. On the other hand, the (return) probability that two 
newly created chain ends recombine immediately must be particulary important at 
large $\omega$. These highly correlated ``diffusion controlled" \protect\cite{oshau} recombination 
events do not contribute do the effective macroscopic reaction rates which determine 
the dynamics of the system. The key task of this paper is to compare different 
methods to compute the effective rates as a function of the attempt frequency $\omega$ 
for a weakly semi-dilute state point in the non-entangled dynamical regime. 
\label{fig0}}
\end{figure}

In the field of self-assembling structures, supramolecular polymers are
attracting nowadays much attention \cite{VdS}. It is well known and schematically 
illustrated in figure \ref{fig0} that some surfactant molecules in solution can 
self-assemble and form wormlike micelles \cite{catc}.

Such micellar solutions exhibit fascinating rheological behaviour,
such as shear-banding\cite{leroug}, shear-thickenning\cite{hoff}, Maxwell fluid behaviour
\cite{catc} or anomalous diffusion (Levy flight)\cite{OBL90}. Self-assembled stacks of discotic 
molecules and chains of bifunctional molecules are other examples of supramolecular polymers. 
All these examples differ by the nature of the intermolecular forces involved in the
self-assembling of the basic units, but they lead to a similar physical
situation bearing much analogy with a traditional system of polydisperse
flexibles polymers when their length becomes sufficiently large with respect to
their persistence length. The specificity and originality of these
supramolecular polymers comes from the fact that these chains are continuously
subject to scissions at random places along their contour and subject to end to
end recombinations, leading to a dynamical equilibrium between different chain
lengths species. These supramolecular polymers are typical soft matter systems
and the chain length distribution which determines their properties, is very
sensitive to external conditions (temperature, concentration, external fields,
salt contents, etc...).

Different times scales and length scales are involved in these systems, as in
usual polymer systems. Unlike the latter, which have been much studied during
the last 50 years\cite{gennes,GK94,doi}, theoretical work on micellar solutions
are quite recent\cite{VdS,catc}. While building up a molecular scale theory is
clearly too heavy for such complex systems, theoretical approaches, based on
mean-field concept and non-local phenomenological approaches,
can explore with success some equilibrium, and rheological
properties\cite{cates,spen,olms}, under reasonable but often drastic
approximations.
Concerning specifically micelle kinetics, recently, O'Shaughnessy and Yu \cite{oshau} 
suggested that there are two
possible kinds of kinetics associated with scission/recombination: diffusion
controlled and mean-field, which may be distinguished by the time dependence of
the first recombination times distribution function of the chains.

In parallel, mesoscopic scale computer simulations can shed much light on these
rich but complex systems, thanks to techniques borrowed from simulations of
polymers. Systems of wormlike micelles can be modelled by ``Equilibrium
polymers" (EP), sometimes called ``living polymers", which are polymer chains
endowed with scission/recombination (S-R) processes taking place in them. The
advantage of numerical studies is their ability to make links either to
experiments or to mean-field type theories, given that those systems are
difficult to characterize by experiments and the results are not simple to
interpret. The bond fluctuation model (BFM) has been the object of intensive
studies on the statics and dynamics of living polymers\cite{witt98,milch}.

Brownian dynamics studies of a similar model \cite{rouau} have been used to
study the scaling predictions on the dependence of the average chain length
upon the overall monomer density. In a recent study, Padding and
Boek\cite{padd} showed that a mesoscopic model of wormlike micelle known as the
FENE-C model\cite{krog}, seems to obey the diffusion-controlled kinetics.

In our work\cite{huang} which is the basis for this course, we investigate the
structural and kinetic properties of equilibrium polymers in solution by means
of a Brownian Dynamics algorithm applied to a standard polymer model\cite{krem} coupled
to a time-reversible EP algorithm for polymer scission-recombinations 
\cite{witt98,milch,MWL99,MW01,WM00}. This particular continuous space model has the
advantage that its kinetic properties are well separated from the structural
and thermodynamical aspects. Our model does not allow for cyclic structures and
hence, we simplify the computational load and, more importantly, the theoretical 
interpretation of the static and dynamical properties of our samples. Anyway, for 
cylindrical micelles, cyclic structures are expected to exist in very low 
concentrations \cite{catc}.

We study the kinetics of equilibrium polymers solutions at two state 
points, exploring in each case a range of rates of the scission-recombination process.
The two state points correspond to a dilute solution and a semi-dilute solution 
regime close to the cross-over region. A direct measure of chain overlap can be 
provided by the ratio of the mean distance between chain centres of mass over the 
radius of giration of the average size chain (respectively $\approx 10$ and $\approx 50$ for the 
dilute and semi-dilute cases). This gives ratios of $1.4$ for the dilute 
case and $0.6$ for the semi-dilute case, which suggests that even in the semi-dilute case, 
the chain relaxation is still typically Rouse like and thus yet kinetically entangled.

The main focus is to link the microscopic model to the more macroscopic kinetics theories, 
e.g. by Cates \cite{catc}, and to discuss in a well defined case the concepts of 
'diffusion controlled' and 'mean-field', defined by authors of\cite{oshau,padd}. In
particular, we want to estimate the effective reaction rates for the scission and 
recombination processes comparing different methods. Furthermore, we shall study 
the diffusion of monomers and discuss their link to the scission/recombination kinetics.

\section{Theoretical framework}

\subsection{Statistical mechanics derivation of the distribution of chain lengths}
To treat a system of supramolecular polymers theoretically, it is convenient to
work at the mesoscopic scale using a model of linear flexible polymers made of
L monomers of size $b$ linked together by a non permanent bonding scheme.
Within the system, individual chain lengths fluctuate by bond scission and by
fusion of two chain ends of different chains. Statistical mechanics can be
employed to predict the equilibrium distribution of chain lengths\cite{catc}.
In terms of the equilibrium chain number density $c_0(L)$, the average chain
length $L_0$ and the total monomer density $\phi$ are given by

\begin{equation}
L_0=\frac{{\sum_{L=0}^{\infty}} L c_0(L)} {\sum_{L=0}^{\infty} c_0(L)}
\label{eq:moy}
\end{equation}
\begin{equation}
\phi=\sum_{L=0}^{\infty} L c_0(L)=\frac{M}{V}\label{eq:mon}
\end{equation}
where $M$ denotes the total number of monomers in the system.

Conceptually, we consider the Helmholtz free energy $F(V,T;\{N(L)\},N_s)$ of a
mixture of chains molecules of different length $L$ in solvent where, in
addition to the temperature T, the volume V and the number of solvent molecules
$N_s$, the number of chains of each specific length $N(L)$ is fixed. Let
$F(V,T;M,N_s)$ be the Helmholtz free energy of a similar system where only the
total number of solute monomers $M$ is fixed. The equilibrium chain length
distribution $c_0(L)=N(L)/V$ will result from the set $\{N(L)\}$ which
satisfies the condition

\begin{equation}
F(V,T;M,N_s)=min_{\{N(L)\}}\left[F(V,T;\{N(L)\},N_s) + \mu \sum_{L=0}^{\infty} L
N(L)\right] \label{eq:fe}
\end{equation}

The parameter $\mu$ is the Lagrange multiplier associated with the constraint
that individual numbers of chains $N(L)$ must keep fixed the total number of
monomers $M=\sum_0^{\infty} L N(L)$. Minimisation requires that the first
derivative with respect to any $N(L')$ variable (L'=1,2, ...) is zero, giving

\begin{equation}
\frac{\delta F(V,T;\{N(L)\},N_s)}{\delta N(L')}+\mu L'=0  ; L'=1,2,...
\end{equation}

We expect the entropic part of the total free energy $F(V,T;\{N(L)\},N_s)$ to
be the sum of translational and chain internal configurational contributions
which both depend upon the way the M monomers are arranged into a particular
chain size distribution. For the translation part, the polydisperse system
entropy is estimated as the ideal mixture entropy $S_{id}$

\begin{equation}
S_{id}(V,T;\{N(L)\},N_s)=-k_B \sum_L N(L) \ln{(C N(L))} + S^{solv}
\label{eq:entro1}
\end{equation}
where $C=b^3/V$ is a dimensionless constant independent of L and where
$S^{solv}$ is the solvent contribution, independent of the $N(L)$
distribution. The configurational entropy of an individual chain with L
monomers is written as $S_1(L)=k_B \ln{\Omega_L}$, in terms of $\Omega_L$,
the total number of configurations of the chain. Adding the configurational
contributions to $S_{id}$ as given by eq.(\ref{eq:entro1}), the total entropy
becomes
\begin{equation}
S(V,T;\{N(L)\},N_s)=-k_B \sum_L N(L) \left[\ln{(C N(L))}-
\ln{\Omega_L}\right] + S^{solv} \label{eq:entro2}
\end{equation}

We now turn to the energy $E(V,T;\{N(L)\},N_s)$ of the same system. If
$E_1(V,T;\\L)$ represents the internal energy of a chain of $L$ monomers and
$E_s$ the energy of a solvent molecule, the energy can be written as
\begin{equation}
E(V,T;\{N(L)\},N_s)=\sum_L N(L) E_1(V,T;L)+N_s E_s(V,T)
\label{eq:energy}
\end{equation}

The key contribution in $E_1$ is the chain end-cap energy $E$ which corresponds
to the chain end energy penalty required to break a chain in two pieces. We
will suppose that $E_1(L)=E+L \tilde{\epsilon}$ where $\tilde{\epsilon}$ is an
irrelevant energy per monomer as $M \tilde{\epsilon}$, its total contribution
to the system energy, is independent of the chain length distribution.

The present approximation of the total free energy of the system is thus given
by incorporating in the general expression (\ref{eq:fe}) the expressions
(\ref{eq:entro2}) and (\ref{eq:energy}), giving
\begin{equation}
\beta F(V,T;\{N(L)\},N_s)=\sum_L N(L)\left[\ln {N(L)} +\ln{C} 
-\ln{\Omega_L}+\beta E +\beta \tilde{\epsilon} L\right]
\end{equation}
where irrelevant constant solvent terms have been omitted as we only need the
first derivative of the free energy with respect to $N(L)$, which now takes the
form
\begin{equation}
\frac{\delta \beta F}{\delta N(L)}= \ln{N(L)} + \ln{C} - \ln{\Omega_L}
+\beta E +\beta \tilde{\epsilon} L + 1.
\end{equation}

With this expression, the minimisation condition on $N(L)$ becomes
\begin{equation}
\ln{N(L)} + \ln{C} - \ln{\Omega_L} + \beta E + 1 +\mu' L = 0 \label{eq:int1}
\end{equation}
where $\mu'=(\beta \mu+\beta \tilde{\epsilon})$. We note at this stage that the
second derivative of $\beta F(V,T;\\\{N(L)\})+\mu' \sum_L L N(L)$ with respect to
$N(L)$ and $N(L')$ variables gives the non negative result
$\frac{\delta_{LL'}}{N(L)}$, indicating that the extremum is indeed a minimum.

Solving for $N(L)$ in eq.(\ref{eq:int1}), we get
\begin{equation}
N(L)={C'}^{-1} \exp{-(\mu' L + \beta E - \ln{\Omega_L})} \label{eq:dist1}
\end{equation}
where $C'=e C$ while, according to eq. (\ref{eq:mon})), $\mu'$ must be such that
\begin{equation}
\sum_L L \exp{-(\mu'L+ \beta E -\ln{\Omega_L})}= M C' \label{eq:norm1}
\end{equation}

The equilibrium $N(L)$ variables are also related to the equilibrium chain
length average $L_0$ (see eq. (\ref{eq:moy})), so that
\begin{equation}
\sum_L \exp{-(\mu'L+ \beta E -\ln{\Omega_L})}= \frac{M C'}{L_0}\label{eq:norm2}
\end{equation}

To progress, we now need to specify the explicit $L$ dependence of $\Omega_L$.
The traditional single chain theories of polymer physics provide universal
expression of $\Omega_L$ in terms of the polymer size, the environment being
simply taken into account through the solvent quality and the swollen blob size 
in the semi-dilute (good solvent) case.

\subsubsection{The case of mean field or ideal chains}

The basic mean-field or ideal chain model for a L segments chain gives

\begin{equation}
\Omega_L^{id}=\left[C_1 z^L\right] \label{eq:ideal}
\end{equation}
where $z$ is the single monomer partition function and $C_1$ a dimensionless
constant. Adapting eq.(\ref{eq:dist1}), one has
\begin{equation}
N(L)=\frac{C_1}{C'}\exp{-(\beta E)} \exp{(-\mu" L)}
\label{eq:dist2}
\end{equation}
where $\mu"=\mu'- \ln{z}$ must, according to eq. (\ref{eq:norm1}), be such that
\begin{equation}
\sum_L L \exp{-(\mu" L)}= \frac{1}{\mu"^2}=\frac{M C'}{C_1 \exp{-(\beta E)}} \label{eq:norm3}
\end{equation}
while eq.(\ref{eq:norm2}) takes the form
\begin{equation}
\sum_L \exp{-(\mu"L)}= \frac{1}{\mu"}=\frac{M C'}{L_0 C_1 \exp{-(\beta E)}}\label{eq:norm4}
\end{equation}

In eqs.(\ref{eq:norm3}) and (\ref{eq:norm4}), sums over $L$ from $1$ to
$\infty$ have been approximated by the result of their continuous integral counterparts.

Combining eqs.(\ref{eq:dist2}), (\ref{eq:norm3}) and (\ref{eq:norm4}), one gets the
final expression for the chain number densities
\begin{equation}
c_0(L)=\frac{\phi}{L_0^2}\exp{(-\frac{L}{L_0})} \label{eq:distsd}
\end{equation}
with the average polymer length given by
\begin{equation}
L_0=B^{1/2} \phi^{\frac{1}{2}} \exp{\left(\frac{\beta E}{2}\right)}.
\label{eq:LL}
\end{equation}
where $B=e b^3/C_1$ is a constant depending upon the monomer size $b$
and the prefactor in the number of ideal chain configurations in eq.(\ref{eq:ideal}).

\subsubsection{The case of dilute chains in good solvent}

Polymer solutions in good solvent are in a dilute regime when chains do not overlap and
in semi-dilute regime when chains do strongly overlap while the total monomer volume fraction 
is still well below its melt value. In the semi-dilute regime, chains remain swollen 
locally over some correlation length, known as the swollen blob size $\chi$, but they
are ideal over larger distances as a result of the screening of excluded volume interactions
between blobs.

Specifically, for a given monomer number density $\phi$, the blob size is given by the condition 
that the blob volume times $\phi$ must be equal to the number of monomers $L^{*}$ in the swollen 
blob. This gives in terms of the reduced number density $\phi'=b^{3}\phi$
\begin{eqnarray}
L^{*}&\sim&{\phi'}^{\left(\frac{1}{1-3\nu}\right)}\label{eq:blo}\\
\chi&\sim&b {\phi'}^{\left(\frac{\nu}{1-3\nu}\right)}
\label{eq:blob}
\end{eqnarray}
where $\nu=0.588$ in present good solvent conditions\cite{doi}.

In living polymers characterized by a monomer number density $\phi$ and some averaged chain length $L_0$, the semi-dilute conditons correspond to the case $L_0>>L^{*}$. We discuss in this subsection the theory for the dilute case where $L_0<<L^{*}$. We will come back to the semi-dilute case in the next subsection.

Self-avoiding walks statistics apply to dilute chains in good solvent, and we thus adopt the number 
of configurations\cite{gennes,GK94}(See especially page 128 of the book of Grosberg and Khokhlov\cite{GK94})
\begin{equation}
\Omega_L^{EV}=C_1 L^{(\gamma-1)} z^L
\label{eq:ev}
\end{equation}
where $\gamma$ is the (entropy related) universal exponent equal to $1.165$.

Incorporating expression (\ref{eq:ev}) in eq. (\ref{eq:dist1}), one gets
\begin{equation}
N(L)=\frac{V}{B}\exp{-(\beta E)}L^{(\gamma-1)} \exp{-(\mu" L)}
\label{eq:dist3}
\end{equation}
where $B$ was introduced in eq.(\ref{eq:LL}) and where $\mu"=\mu'-\ln{z}$ must be fixed by eq.(\ref{eq:mon})
\begin{equation}
\sum_L L^{\gamma} \exp{-(\mu" L)}=B \phi \exp{(\beta E)} \label{eq:norm5}
\end{equation}
while eq.(\ref{eq:norm2}) takes here the form
\begin{equation}
\sum_L L^{(\gamma-1)}\exp{-(\mu"L)}=\frac{B \phi}{L_0} \exp{(\beta E)}\label{eq:norm6}
\end{equation}

If $L$ is treated as a continuous variable, eqs(\ref{eq:norm5}) and (\ref{eq:norm6}) can be rewritten in terms of the Euler Gamma function satisfying $\Gamma(x)=x \Gamma(x-1)$ as
\begin{equation}
\int_0^{\infty} L^{\gamma} \exp{-(\mu" L)}dL = \frac{\Gamma(\gamma+1)}{\mu"^{(\gamma+1)}} = B \phi \exp{(\beta E)} \label{eq:norm7}
\end{equation}
\begin{equation}
\int_0^{\infty}  L^{(\gamma-1)}\exp{-(\mu"L)}dL= \frac{\Gamma(\gamma)}{\mu"^{\gamma}}=\frac{B \phi}{L_0} \exp{(\beta E)}\label{eq:norm8}
\end{equation}
From eqs (\ref{eq:norm7}) and (\ref{eq:norm8}), one gets
\begin{eqnarray}
\mu"&=&\frac{\gamma}{L_0}\\
B \phi \exp{(\beta E)}&=&\frac{\Gamma(\gamma+1)}{\gamma^{(\gamma+1)}} L_0^{(\gamma+1)}
\end{eqnarray}
These results lead then finally to the Schulz-Zimm distribution of chain lengths, namely
\begin{equation}
c_0(L)=\frac{\exp{(-\beta E)}}{B}L^{(\gamma-1)} \exp{(-\gamma \frac{L}{L_0})}
\label{eq:distdi}
\end{equation}
and an average polymer length given by
\begin{equation}
L_0=\left(\frac{\gamma^{\gamma}}{\Gamma(\gamma)}\right)^{\frac{1}{1+\gamma}}B^{\frac{1}{1+\gamma}}
\phi^{\frac{1}{1+\gamma}} \exp{\left(\frac{\beta E}{(1+\gamma)}\right)}.
\end{equation}

\subsubsection{The semi-dilute case}

We consider here the semi-dilute case in good solvent where the average length of living polymers $L_0$ is much larger than the blob length $L^{*}$.
The usual picture of a semi-dilute polymer solution is an assembly of ideal chains made of blobs of size $\chi$. Using this approach, Cates and Candau \cite{catc} and later J.P. Wittmer et al \cite{witt98} derived the relevant equilibrium polymer size distribution. In this subsection, we adapt their derivation to the theoretical framework presented above. 

Let $\Omega_b$ be the number of internal configurations per blob and $z'$ some coordination number for successive blobs. As there are $n_b=L/L^{*}$ blobs for a chain of L monomers, we write the total number of internal configurations of a chain of size L as
\begin{equation}
\Omega_L^{SD}=C_1 L^{*(\gamma-1)} \Omega_b^{L/{L^{*}}} {z'}^{L/{L^{*}}}
\label{eq:sd}
\end{equation}
where $\gamma$ is the universal exponent in the excluded volume chain statistics met earlier for chains in dilute solutions. The important factor $L^{*(\gamma-1)}$ can be seen as an entropy correction for chain ends just like $E$ was an energy correction to $L\tilde{\epsilon}$. This entropic term which involves the number of monomers per blob, is needed to take into account that when a chain breaks, its two ends are subject to a reduced excluded volume repulsion. The other factors in eq.(\ref{eq:sd}) will lead to terms linear in $L$ after taking the logarithm and thus will be absorbed in the Lagrange multiplier definition, as seen earlier in similar cases for ideal and dilute chains.The resulting expression of $N(L)$ in terms of the Lagrange multiplier (cfr eq.(\ref{eq:dist2})) can then be written by analogy as
\begin{equation}
N(L)=\frac{C_1}{C'}\exp{-\left[\beta E-(\gamma-1)\ln{L^{*}}\right]} \exp{(-\mu L)}
\end{equation}

Proceeding as in the ideal case (simply replacing at every step the constant $\beta E$ by $\beta E-(\gamma-1)\ln{L^{*}}$, one recovers in the semi-dilute case the simple exponential distribution 
\begin{equation}
c_0(L)=\frac{\phi}{L_0^2}\exp{(-\frac{L}{L_0})}
\label{eq:dist4}
\end{equation}
with a slightly different formula for the average polymer length
\begin{equation}
L_0 \propto \phi^{\alpha} \exp{\left(\frac{\beta E}{2}\right)}
\end{equation}
where $\alpha=\frac{1}{2}(1+\frac{\gamma-1}{3\nu-1})$ is about 0.6.

\subsection{A kinetic model for scissions and recombinations}

The interest for wormlike micelles dynamics came from the experimental
observation that entangled flexible supramolecular polymers display, after
an initial strain, a simple exponential stress relaxation which is
qualitatively different from the behaviour of usual entangled melts. In the
latter system, individual chains must leave by a reptation mechanism the strained
topological tube created by the entangled temporarily network, in order to relax the 
shearing forces. A plausible scission-recombination model, allowing for such an extra
relaxation mechanism, was shown \cite{catc} to lead to an exponential decay
of the shearing forces with a decay Maxwell time equal to
$\tau \approx \sqrt{\tau_b \tau_{rep}}$ where $\tau_{rep}$ is the chain 
reptation time and $\tau_b$ is the mean life time of a chain of average size 
in the system. 

We will assume in the following that the Cate's scission-recombination model governing 
the population dynamics, originally deviced to explain entangled equilibrium polymer 
melt rheology, should also apply to the kinetically unentangled regime which is explored in 
the present work.

This kinetic model \cite{cates} assumes that
\begin{itemize}
\item  the scission of a chain is a unimolecular process, which occurs with
equal probability per unit time and per unit length on all chains. The rate of
this reaction is a constant $k_s$ for each chemical bond, giving
\begin{equation}
\tau_b=\frac{1}{k_s L_0}
\label{taub}
\end{equation}
for the lifetime of a chain of mean length $L_0$ before it breaks into two
pieces.
\item recombination is a bimolecular process, with a rate $k_r$ which is identical
for all chain ends, independent of the molecular weight of the two reacting
species they belong to. It is assumed that recombination takes place with a new
partner respect to its previous dissociation as chain end spatial correlations are 
neglected within the present mean field theory approach. It results from detailed 
balance that the mean life time of a chain end is also equal to $\tau_b$.
\end{itemize}

Let $c(t,L)$ be the number of chains per unit volume having a size $L$ at time
$t$. Cates \cite{cates} writes the kinetic equations as
\begin{eqnarray}
\frac{d c(t,L)}{dt}&=&-k_s L c(t,L)+ 2k_s \int_L^{\infty} c(t,L') {dL'} \\
&+&\frac{k_r}{2}\int_0^{L} c(t,L')c(t,L-L')dL' -k_r c(t,L)\int_0^{\infty} c(t,L'){dL'}
\end{eqnarray}
where the two first terms deal with chain scission (respectively disappearence or appearence
of chains with length $L$) while the two latter terms deal with chain
recombination (respectively provoking the appearence or disappearence of chains of
length $L$).

It is remarkable that the solution of this empirical kinetic model leads to an
exponential distribution of chain lengths. Indeed, direct substitution of
solution $c_0(L)$ in the above equation leads to the detailed balance
condition:
\begin{equation}
\phi \frac{k_r}{2k_s}=L_0^2
\end{equation}
the ratio of the two kinetic constant being thus restricted by the
thermodynamic state. Detailed balance means that for the equilibrium
distribution $c_0(L)$, the number of scissions is equal to the number of
recombinations. The total number of scissions and recombinations per unit
volume and per unit time, denoted respectively as $n_s$ and $n_r$, can be
expressed as
\begin{eqnarray}
n_s&=& k_s \frac{\phi}{L_0^2} \int_0^{\infty}L \exp{(-L/L_0)}dL =k_s \phi \label{eq:ntran1}\\
n_r&=& \frac{k_r}{2} \frac{\phi^2}{L_0^4} \int_0^{\infty} dL'\int_0^{\infty} dL"
\exp{(-\frac{L'}{L_0})}  \exp{(-\frac{L"}{L_0})}= \frac{\phi^2 k_r}{2 L_0^2}
\label{eq:ntran2}
\end{eqnarray}
and it can be easily verified that detailed balance condition implies
$n_s=n_r$.

Mean field theory assumes that a polymer of length $L$ will break on average
after a time equal to $\tau_b=(k_s L)^{-1}$ through a Poisson process. This
implies that the distribution of first breaking times (equal to the survival
times distribution) must be of the form
\begin{equation}
\Psi(t) = \exp{(-\frac{t}{\tau_b})} \label{taub}
\end{equation}
for a chain of average size. Detailed balance then requires that the same
distribution represents the distribution of first recombination for a chain end\cite{catc}.
Accordingly, throughout the rest of this chapter, the symbol $\tau_b$ will represent 
as well the average time to break a polymer of average size or the average time between
end chain recombinations. In the same spirit, we stress that among the different estimates 
of $\tau_b$ proposed in this work, some are based on analyzing the scission statistics while others 
are based on the recombination statistics. 

Two additional points may be stressed at this stage:
\begin{itemize}
\item The mean field model in the present context has been questionned \cite{oshau} because in many
applications, there are indications that a newly created chain end often
recombines after a short diffusive walk with its original partner. In that
case, a possibly large number of breaking events are just not effective and the kinetics
proceeds thus differently.
\item Given the statistical mechanics analysis in the previous subsection, we see that
the equilibrium distribution of chain lengths resulting from the simple empirical kinetic 
model is perfectly compatible with the equilibrium distribution in polymer solutions at the 
$\theta$ point (ideal chains) or for semi-solutions (ideal chains of swollen blobs). The 
kinetic model is still pertinent, given its simplicity, in the case of dilute solutions 
as the static chain length distribution based on statistical mechanics, although non exponential, 
is not very far from it.
\end{itemize}

\section{The mesoscopic model and its Brownian Dynamics implementation\label{seco2}}

In order to test the theoretical understanding summarized in the previous section, we have 
devised a new microscopic model bearing resemblances with an off-lattice
Monte-Carlo model proposed earlier \cite{MWL99}, but which uses Brownian Dynamics 
to follow the monomer dynamics for a standard polymer chain model\cite{krem}. 
To be able to clearly unravel static and dynamic aspects, our model contains a kinetic 
parameter which allows to vary the rates of the reactions without affecting the static and
thermodynamic properties. As already mentioned in the introduction, two 
thermodynamic states corresponding to equilibrium polymers dissolved in a good 
solvent are considered, namely a dilute and a semi-dilute solution.

We consider a set of micelles consisting of (non-cyclic) linear assemblies of
Brownian particules. Within such a linear assembly, the bounding potential
$U_1(r)$ acting between adjacent particles is expressed as the sum of a
repulsive Lennard-Jones (shifted and truncated at its minimum) and an
attractive part of the FENE type\cite{krem}.
The pair potential $U_2(r)$ governing the interactions between any unbounded pair 
(both intramicellar and intermicellar) is a pure repulsive potential corresponding 
to a simple Lennard-Jones potential shifted and truncated at its minimum. This 
choice of effective interaction between monomers implies good solvent conditions.

\begin{figure}[th] 
\centerline{\includegraphics[width=.75\textwidth,angle=-90]{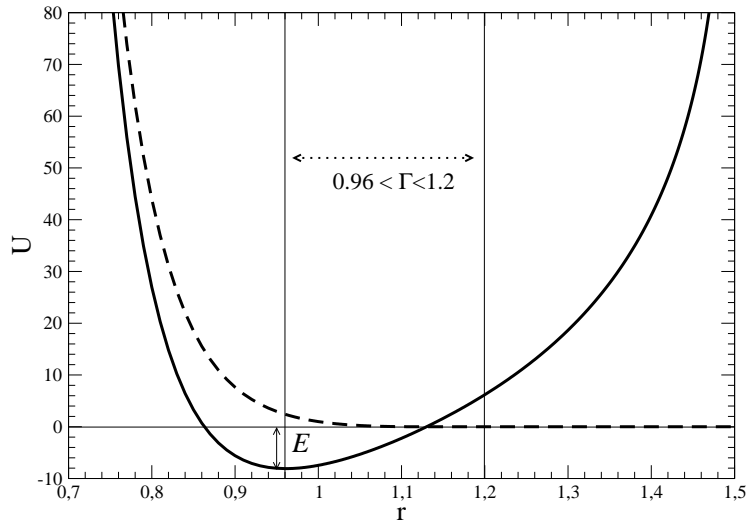}}
\vspace*{8pt}\caption{Bounded potential $U_1(r)$ (continuous curve) and unbounded 
potential $U_2(r)$ (dashed line) between a pair of monomers. Each monomer can participate 
to at most two bounded links (with potential $U_1(r)$). All unbounded monomer pairs 
interact via the potential $U_2(r)$. E is a parameter tuning the energy required 
to open the bond. The figure also shows the $\Gamma$ region where potential swaps (equivalent 
to bond scissions or bond recombinations) are allowed during the brownian dynamics simulation is 
indicated.\label{fig1}}
\end{figure}

Using the Heaviside function $\Theta(x)=0$ or $1$ for $x<0$ or $x\geq 0$
respectively, explicit expressions (see figure~\ref{fig1}) are
\begin{eqnarray}
U_2(r)&=& 4\epsilon \left[(\frac{\sigma}{r})^{12}-(\frac{\sigma}{r})^{6}+
\frac{1}{4}\right] \Theta(2^{1/6}\sigma-r))\\
U_1(r)&=& U_2(r)- 0.5 k R^2 \ln{\left(1-[\frac{r}{R}]^2\right)}-U_{min}-E
\label{eq:pot}
\end{eqnarray}

In the second expression valid for $r<R$, $k=30 \epsilon/\sigma^2$ is
the spring constant and $R=1.5\sigma$ is the value at which the FENE potential
diverges. $U_{min}$ is the minimum value of the sum of the two first terms of
the second expression (occuring $r_{min}=0.96094\sigma$ for the adopted
parameters) while $E$ is a key parameter which corresponds, when pair potential
exchanges between bounded and unbounded situations will be allowed, to the
typical energy gain (loss) when an unbounded (bounded) pair is the object of a
recombination (scission).

We deal with M monomers at temperature $T$ enclosed in a volume V, defining a
monomer number concentration $\phi=M/V$. The relevant canonical partition
function is a sum over all microscopic states specified not only by monomer
positions (and momenta) but also by a set of variables defining unambiguously 
the connectivity scheme for the particular microscopic state.
Disregarding the irrelevant momentum variables in Brownian Dynamics, we write the 
partition function as

\begin{equation}
Q_{MVT} \sim \sum_{\lbrace r \rbrace} \sum_{m}\exp{(-\frac {U(\lbrace r
\rbrace,m)}{k_B T})}
\end{equation}
where the double sum runs over all possible arrangements of M distinguishable 
monomers and over all related bounding schemes. The first sum runs over all spatial 
positions $\lbrace r\rbrace$ of the M monomers, each configuration being characterized 
by a list of K monomer-monomer pairs with relative distance shorter than R. The second sum
runs, for a given spatial arrangement with a subset of K pairs available for
bounding, over all allowed distinct bounding schemes, namely those among the
$2^{K}$ distinct possibilities which satisfy two restricting rules, namely
\begin{itemize}
\item no monomer can be engaged into more than two bounding pairs (= no branching).
\item no cyclic bounding structure is allowed.
\end{itemize}

The potential energy associated to a particular configuration can be written
\begin{equation}
U(\lbrace r \rbrace,m)=\frac{1}{2} \sum_{i}{\sum_{j\neq i} {(s_{ij}
U_1(r_{ij})+(1-s_{ij}) U_2(r_{ij}))}}
\end{equation}
where $s_{ij}=0$ for all pairs with $r_{ij}>R$ while $s_{ij}=$ either 0 or 1
for the K pairs with $r_{ij}<R$. Index m in the two previous expressions
corresponds to a particular bounding network representing a particular set of 0
and 1 for the K $s_{ij}$ values.

In order to discuss topological changes (change in bounding network) during the
phase space exploration, it is useful to consider that each monomer possesses
two arms available for bounding. If both arms are free, we have an isolated
monomer with degree of polymerisation $L=1$. If at least one arm of a
particular monomer is linked through potential $U_1(r)$ to an arm of another
monomer, these two monomers belong to the same linear structure (micelle) of
degree of polymerisation $L\geq2$. A terminal or an interior monomer will
correspond to a monomer engaged respectively in one or two bonds with other
monomers.

The phase space exploration will be performed by a Brownian Dynamics (BD) scheme
(controling $r$ positions updates) coupled to a single bond creation
(replacement of potential $U_2(r)$ by $U_1(r)$) / single bond annihilation
(replacement of potential $U_1(r)$ by $U_2(r)$) scheme satifying microscopic
reversibility.

The BD is standard and samples the set of positions of the M monomers according
to a canonical ensemble corresponding to a particular potential energy
(equivalent in the present case to a particular bonding network m).

The single BD step of particule i subject to a total force $\overline{F}_i$ is
simply
\begin{equation}
\overline{r}_i(t + \Delta t)= \overline{r}_i(t) +\frac{\overline{F}_i}{\xi}
\Delta t+ \overline{R}_i(\Delta t)
\end{equation}
where the last term corresponds to a vectorial random gaussian quantity with
first and second moments given by
\begin{eqnarray}
<{R}_{i \alpha}(\Delta t)>&=&0 \\
<{R}_{i \alpha}(\Delta t) {R}_{j \beta}(\Delta t)>&=& 2 (\frac{k_B T}{\xi})
\Delta t \delta_{\alpha \beta} \delta_{ij}
\end{eqnarray}
for arbitrary particles i and j and where $\alpha \beta$ stand for the
$x,y,z$ cartesian components. 

At this stage, it is useful to fix units. In the following, we will adopt the size 
of the monomer $\sigma$ as unit of length, the $\epsilon$ parameter as energy unit 
and we will adopt $\xi\sigma^{2}/(3 \pi \epsilon)$ as time unit. We also introduce the reduced
temperature $k_B T/\epsilon=T^{*}$,  so that in reduced coordinates (written with symbol*)
the algorithm becomes
\begin{eqnarray}
<{R}_{i \alpha}^{*}(\Delta t^{*})>&=&0 \\
<{R}_{i \alpha}^{*}(\Delta t^{*}) {R}_{j \beta}^{*}(\Delta t^{*})>&=& \frac{2}{3 \pi} \Delta
t^{*} \delta_{\alpha \beta} \delta_{ij} \,T^{*}
\end{eqnarray}
which means that each particle, if isolated in the solvent, would diffuse with
a RMSD of $\sqrt{2 T^{*}/\pi}$ per unit of time.

The bonding network is itself the object of random instantaneous changes
provided by a Monte-Carlo algorithm which is built according to the standard
Metropolis scheme.

The probability $P_{m,n}$ to go from a bounding network $m$ to a different one
$n$ is written as
\begin{equation}
P_{m,n}= A_{m,n}^{trial} P_{m,n}^{acc}
\end{equation}
where $A_{m,n}^{trial}$ is the trial probability to reach a new bounding
network $n$ starting from the old one $m$, within a single MC step. This trial
probability is chosen here to be symmetric as usually adopted in Metropolis
Monte Carlo schemes. This trial probability is chosen to be different from zero
only if both bounding networks n and m differ by the status of a single bond,
say the pair of monomers $(ij)$. To satisfy microreversibilty, the acceptance
probability $P_{m,n}^{acc}$ for the trial ($m \longrightarrow n$) must be given
by
\begin{equation}
P_{m,n}^{acc}= Min[1,\exp{(-\frac{(U(\lbrace r \rbrace,n)-U(\lbrace r
\rbrace,m))}{k_B T})}].
\end{equation}

In the present case, as a single pair (ij) changes its status, the acceptance
probability takes the explicit form
\begin{eqnarray}
P_{m,n}^{acc}&=&Min[1,\exp{(-\frac{(U_2(r_{ij})-U_1(r_{ij}))}{k_B T})}] \\
P_{m,n}^{acc}&=&Min[1,\exp{(-\frac{(U_1(r_{ij})-U_2(r_{ij}))}{k_B T})}]
\end{eqnarray}
for bond scission and bond recombination respectively.

\begin{figure}[th] 
\centerline{\includegraphics[width=.75\textwidth]{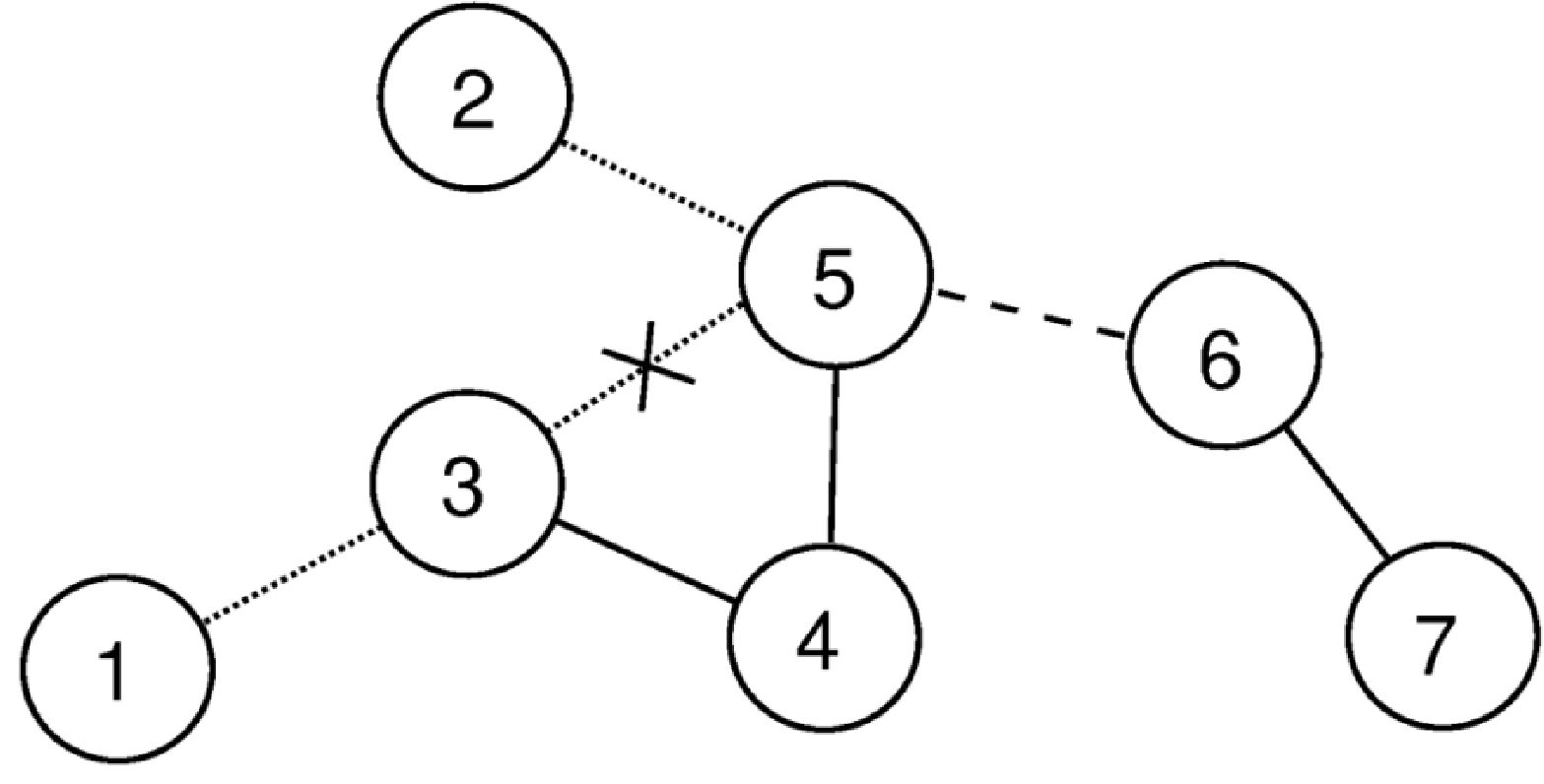}}
\vspace*{8pt}\caption{Exemplary configuration of a 7 monomers system in state n where monomers 
3,4 and 5 form a trimer and monomers 6 and 7 a dimer. All pairs of monomers with 
mutual distances within the $\Gamma$ region are indicated by a dotted or a dashed line.
In the text, we consider the Monte Carlo scheme for transitions between states n and m 
which only differ by the fact that in state n and m the 5-6 pair is respectively open 
or bounded. The n $\rightarrow$ m transition corresponds to the creation 
of a pentamer by connecting a dimer and a trimer while the m $\rightarrow$ n transition 
leads to the opposite scission. The cross symbol on link 3-5 indicates that in state n, when 
looking to all monomers which could form a new link with monomer 5, monomer 3 is excluded 
because it would lead to a cyclic polymer which is not allowed within the present model. \label{fig2}}
\end{figure}

The way to specify the trial matrix $A_{m,n}^{trial}$ starts by defining a
range of distances, called $\Gamma$ and defined by $0.96<r<1.20$ within the
range $r<R$. For $r \in \Gamma$, a change of bounding is allowed as long as the
two restricting rules stated above are respected. Consider the particular
configuration illustrated by figure \ref{fig2} where M=7 monomers located at the shown
positions, are characterized by a connecting scheme made explicit by
representing a bounding potential by a continuous line. The dashed line between
monomers 5 and 6 represents the changing pair with distance $r \in \Gamma$ which
is a bond $U_1(r)$ in configuration $m$ but is just an ordinary intermolecular
pair $U_2(r)$ in configuration $n$.

For further purposes, we have also indicated in figure \ref{fig2} by a dotted line all
pairs with $r \in \Gamma$ which are potentially able to undergo a change from a
non bounded state to a bounded one in the case where the (56) pair, on which we
focus, is non bounded (state n). Note that bond (35) is not represented by a
dotted line eventhough the distance is within the $\Gamma$ range: a bond
formation in that case is not allowed as it would lead to a cyclic
conformation. Note also that in state n, monomer 5 could thus form bonds either
with monomer 2 or with monomer 6 while monomer 6 can only form a bond with
monomer 5.

We now state the algorithm and come back later on the special
($m\longrightarrow n$) transition illustrated in figure \ref{fig2}.

During the BD dynamics, with an attempt frequency $\omega$ per arm and per unit
of time, a change of the chosen arm status (bounded $U_1(r)$ to unbounded
$U_2(r)$ or unbounded to bounded) is tried. If it is accepted as a ``trial move"
of the bounding network, it obviously implies the modification of the status of
a paired arm belonging to another monomer situated at a distance $r \in \Gamma$
from the monomer chosen in the first place.

The trial move goes as follows: a particular arm is chosen, say arm 1 of
monomer i, and one first checks whether this arm is engaged in a bounding pair
or not.
\begin{itemize}
\item If the selected arm is bounded to another arm (say arm 2 of monomer j) and the
distance between the two monomers lies within the interval $r_{ij}\in\Gamma$, an opening
 is attempted with a probability $1/(N_i+1)$, where the integer $N_i$ represents
 the number of monomers available for bonding with monomer i, besides the monomer j
 ($N_i$ is thus the number of monomers with at least one free arm whose distance to the
  monomer carrying the originally selected arm i lies within the interval $\Gamma$,
  excluding from counting the monomer j and any particular arm leading to a ring
  closure). If the trial change consisting in opening the (ij) pair is refused
  (either because the distance is not within the $\Gamma$ range or because the
opening attempt has failed in the case $N_i>0$), the MC step is stopped without
bonding network change (This implies that the BD restarts with the (ij) pair
being bonded as before).
\item If the selected arm (again arm 1 of monomer i) is free from bonding, a search
 is made to detect all monomers with at least one arm free which lie in the ``reactive"
 distance range $r\in \Gamma$ from the selected monomer i (Note that if monomer i is a
 terminal monomer of a chain, one needs to eliminate from the list if needed, the other
  terminal monomer of the same chain in order to avoid cyclic micelles configurations).
  Among the monomers of this ``reactive" neighbour list, one monomer is then selected at
  random with equal probability to provide an explicit trial bonding attempt between
  monomer i and the particular monomer chosen from the list. Note that if the list is
  empty, it means that the trial attempt to create a new bond involving arm 1 of monomer
  i has failed and no change in the bonding network will take place.
\end{itemize}

In both cases, if a trial change is proposed, it will be accepted with the
probability $P_{m,n}^{acc}$ defined earlier. If the change is accepted, BD will
be pursued with the new bonding scheme (state n) while if the trial move is
finally rejected, BD restarts with the original bonding scheme corresponfing to
state m.

Coming back to figure \ref{fig2}, we now show that the MC algorithm mentioned above
garantees that the matrix $A_{mn}^{trial}$ is symmetric, an important issue as
it leads to the micro-reversibility property when combined with the acceptance
probabilities described earlier. Let us define as $P_{arm}=1/2N$ the
probability to select a particular arm, a uniform quantity.

If configuration m with pair (56) being ``bounded" is taken as the starting
configuration, the number of available arms to form alternative bonds with
monomer 5 and monomer 6 are respectively $N_5=1$ and $N_6=0$. Therefore,
applying the MC rules described above, the probability to get configuration n
where the pair (56) has to be unbounded is given by the sum of probabilities to
arrive at this situation through selection of the arm of monomer 5 engaged in
the bond with monomer 6 or through selection of the arm of monomer 6 engaged in
a bond with monomer 5. This gives
\begin{equation}
A_{m,n}^{trial}=P_{arm}*\frac{1}{N_5+1}+P_{arm}\frac{1}{N_6+1}=\frac{3}{2}*P_{arm}
\end{equation}

If configuration n with pair (56) being ``unbounded" is taken as the starting
configuration, the application of the MC rules lead to the probability to get
configuration m where the pair (56) has to be bounded is given by the sum of
probabilities to arrive at this situation through selection of the free arm of
monomer 5 (which has two bounding possibilities, namely with monomers 2 and 6)
or through selection of the free arm of monomer 6 which can only form a bond
with monomer 5. This gives
\begin{equation}
A_{n,m}^{trial}=P_{arm}*\frac{1}{2}+P_{arm}=\frac{3}{2}*P_{arm}
\end{equation}
showing the required matrix symmetry.

\section{Characterization of static properties}

In the following, all quantities will systematically be expressed in reduced units without 
an explicit $*$ symbol.

\subsection{List of experiments and chain length distributions}

We consider a system of $M=1000$ monomers interacting via the potential model
defined by eqs.(\ref{eq:pot}) at $T=1$. This is a rather small system by nowadays standards 
but we were only interested here to a preliminary demonstrative study before going in the future 
to larger systems and a more efficient description of solvent effects. 

This system is studied at two state points:
\begin{itemize}
\item A dilute solution at the number density $\phi=0.05$ and an energy
parameter $E=8$.
\item A semi-dilute solution at the number density $\phi=0.15$
and an energy parameter $E=10$.
\end{itemize}
Each system evolves according to the Brownian Dynamics algorithm with time step
$\Delta t=0.001$ and is subject to random trials of bond scission/recombination 
with arm trial attempts frequencies in the range $0.1<\omega <5$. This choice 
was motivated by the necessity to speed up the kinetics and hence, 
getting reasonably good statistics on dynamical properties. Obviously, all static properties 
are independent of $\omega$ and therefore, all data can be cumulated over all $\omega$ values.

The systems have been equilibrated for a time of $2\ 10^{4}$. Production
runs have been followed for a similar time $T_{run}=2\ 10^{4}$. The equilibrium
values of the mean chain length $L_0$ and the total number of chains $<N_{ch}>$
are related by \linebreak$M=L_0<N_{ch}>$. 


\subsubsection{The dilute case}

The first experiment deals with a dilute solution as the average chain length
$L_0=10.4 \pm0.1$ is much smaller than the crossover value at that monomer number
density as calculated by eq.(\ref{eq:blob}), $L^{*}=50.5$.

\begin{figure}[th] 
\centerline{\includegraphics[width=.75\textwidth,angle=-90]{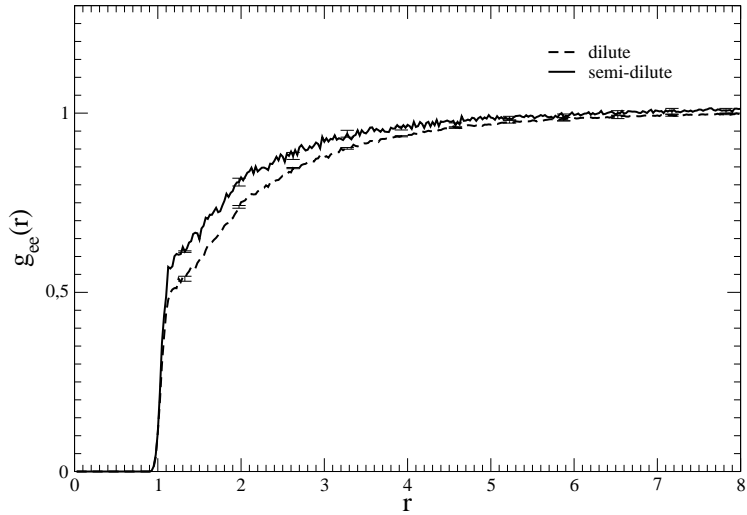}}
\vspace*{8pt}\caption{Chain end pair distribution function $g_{ee}(r)$ in the dilute 
and semi-dilute experiments. Note that the $\Gamma$ region where the bounding changes
take place correspond to the region of first (fast) increase of the distribution 
function around $r=\sigma$. \label{fig3a}}
\end{figure}

\begin{figure}[th] 
\centerline{\includegraphics[width=.75\textwidth,angle=-90]{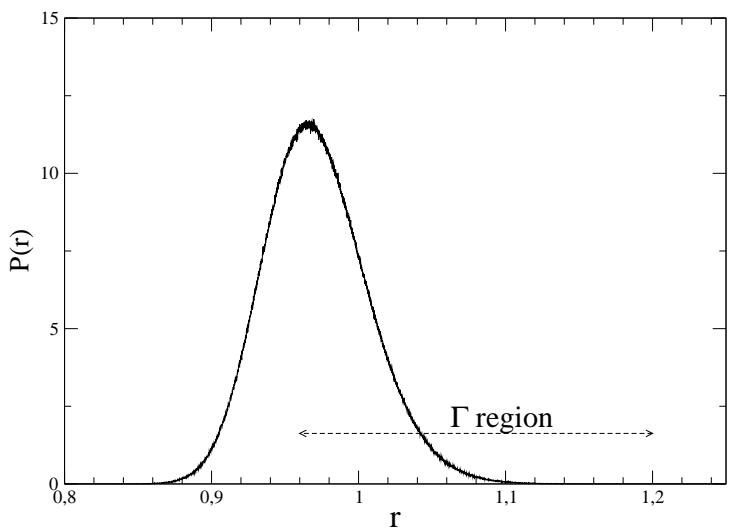}}
\vspace*{8pt}\caption{Normalized distribution function $P(r)$ of the distance between bounded monomers 
in the semi-dilute experiment. The corresponding function in the dilute solution experiment 
is marginally different from the semi-dilute case and is therefore not shown in the figure.
\label{fig3b}}
\end{figure}

The populations of bonds ready to open within the $\Gamma$ range is
$<N_{1\Gamma}>=551.6$ and represents $61 \%$ of the bonded pairs $<N_1>$, while
the population of free arms pairs ready to close, again within the same
$\Gamma$ range, is only $<N_{2\Gamma}>=1.16$. Figures~\ref{fig3a} and
~\ref{fig3b} show respectively the pair correlation function for chain ends
(unsaturated arms) and the distribution of the distance between bounded
monomers, in particular within the $\Gamma$ region where potential changes do
occur. The ``free arm" fraction, namely $1-N_1/M=0.0957$, is close to
$L_0^{-1}=<N_{ch}/M>=0.0964$.

\begin{figure}[th] 
\centerline{\includegraphics[width=.75\textwidth,angle=-90]{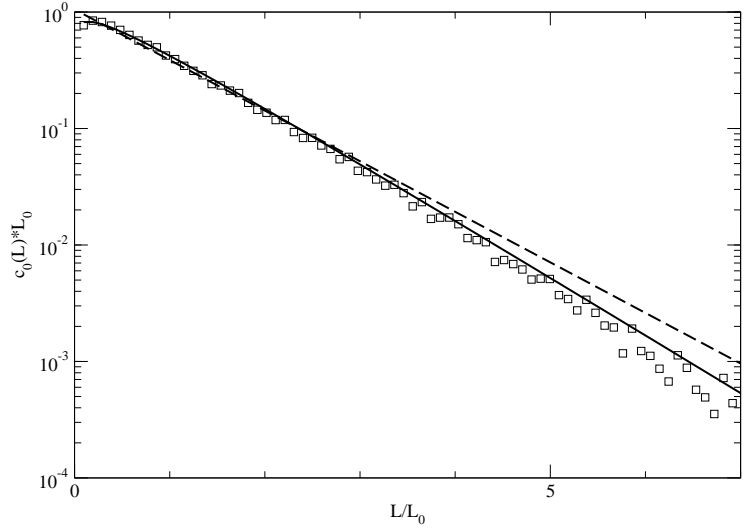}}
\vspace*{8pt}\caption{Distribution of chain length number densities for the dilute case in 
good solvent. The data (squares) are fitted using eq.(\ref{eq:distdi}) with imposed value 
$\gamma=1.165$. The dashed line shows the exponential distribution $C_0(L) \propto 
\exp{(-\frac{L}{L_0})}$ which does not fit data as well.
\label{fig4a}}
\end{figure}

Dilute solution conditions are confirmed by the chain length distribution shown
in figure~\ref{fig4a}. For dilute conditions, a distribution given by
eq.(\ref{eq:distdi}) is expected. Accordingly, we have fitted our data with a
single parameter (prefactor) fit function, $c(L)=A_0 L^{\gamma-1} \exp{(-\gamma L/L_0)}$ where
$L_0$ was given its computed average value and where $\gamma$ was given its expected 
value, $\gamma=1.165$. This curve is significantly better than the simple exponential distribution
expected for ideal or semi dilute chains. If $\gamma$ is left as a second free parameter in the fit, 
it takes an even larger value $1.187$ and the fit (not shown) is slightly better especially at high L.

\subsubsection{The semi-dilute case}

In the second state point experiments, the average chain length is found to be \linebreak$<L>=49.9 \pm0.1$, a
value which is much larger than the crossover value $L^{*} \approx 12$ at $\phi=0.15$ according to 
eq.(\ref{eq:blo}).

The populations of bonds ready to open
in the $\Gamma$ region is $<N_{1\Gamma}>=597 \pm 1.$, while the population of
free arms pairs ready to close is $<N_{2\Gamma}>=0.183$. Distance distributions
of unbounded and bounded pairs are given in figures \ref{fig3a} and \ref{fig3b} together with the
dilute case data. We note that $g(r)$ function between free ends is not very
different between the two cases. The ``free arm" fraction $1-N_1/M=0.021$ is 5
times smaller than in the dilute case as the chains are 5 times longer.

\begin{figure}[th] 
\centerline{\includegraphics[width=.75\textwidth,angle=-90]{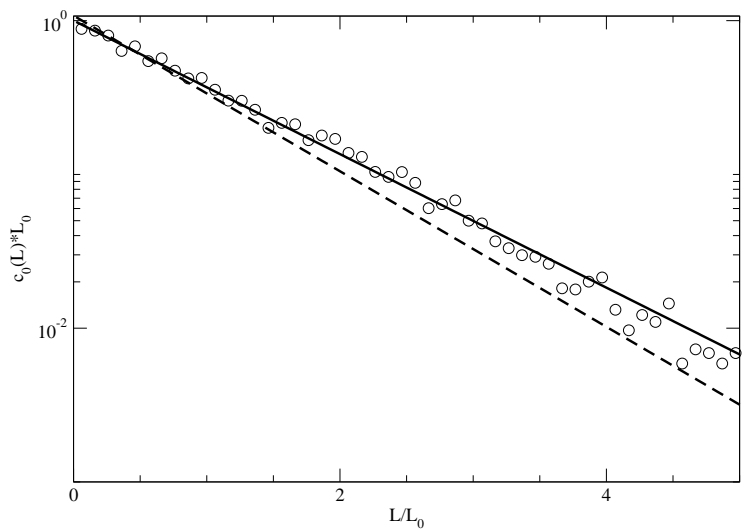}}
\vspace*{8pt}\caption{Distribution of chain lengths number densities for the 
semi-dilute case. The data (circles) are fitted with a simple exponential function 
$A \exp{(-L/L_0)}$ (continuous curve), in reference to eq.(\ref{eq:dist4}) with imposed 
average chain size $L_0=49.9$. Dashed lines show a function $\exp{(-1.165 \frac{L}{L_0})}$ 
which does not fit the data as well. \label{fig4b}}
\end{figure}

Semi-dilute solution conditions are confirmed by the observation of an
exponential distribution of the reduced chain length (see figure~\ref{fig4b}). In the 
latter figure, we show the best fit by a one parameter fitting function 
$c^{fit}(L)=A \exp{(-L/L_0)}$.

\subsection{Chain length conformational analysis}

\begin{figure}[th] 
\centerline{\includegraphics[width=.75\textwidth,angle=-90]{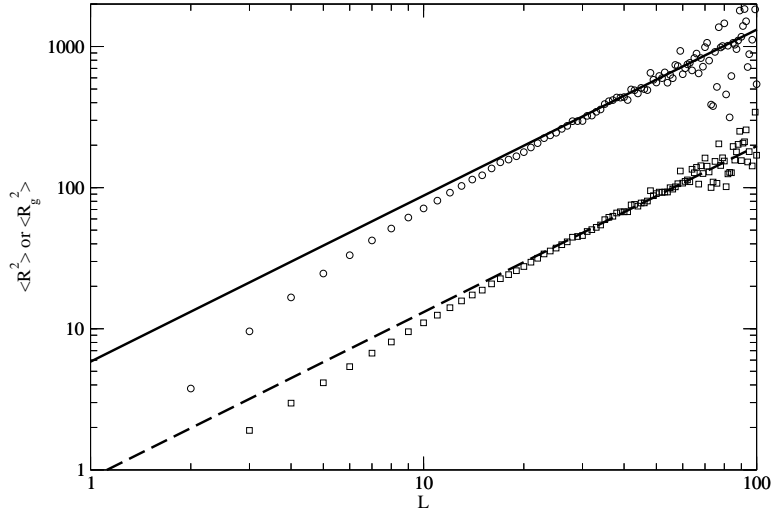}}
\vspace*{8pt}\caption{Conformational properties in the dilute case: $<R^{2}>$ (circles) and $<R_{g}^{2}>$ (squares)
versus chain length $L$. The fitting functions which are power law with
exponent fixed to $2\nu=1.176$ (prefactors $0.876$ for $<R_g^{2}>$ and $5.854$ 
for $<R^{2}>$) fit the data for longer chains only \label{fig5a}}
\end{figure}

\begin{figure}[th] 
\centerline{\includegraphics[width=.75\textwidth,angle=-90]{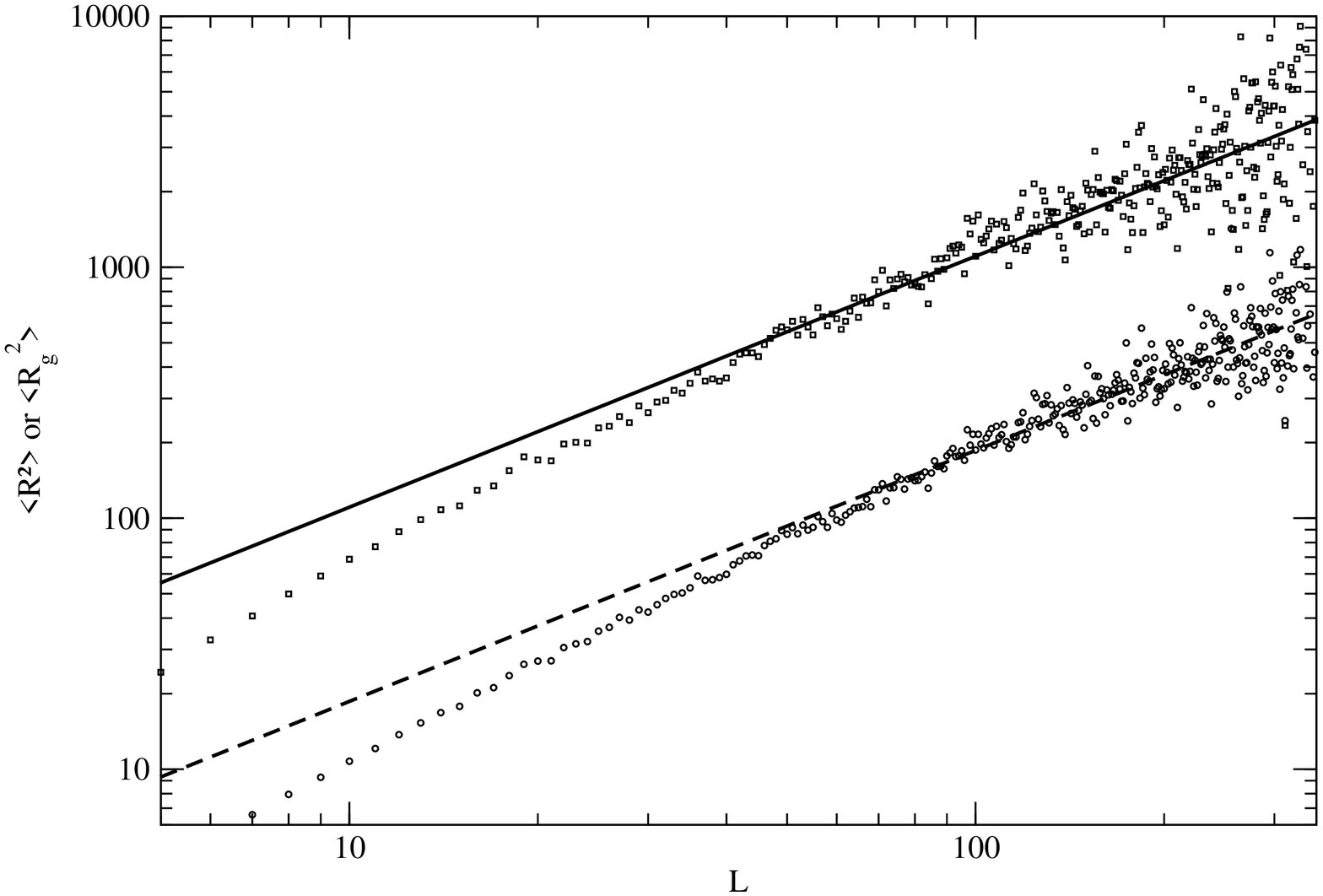}}
\vspace*{8pt}\caption{Conformational properties in the semi-dilute case: $<R^{2}>$ (squares) 
and $<R_{g}^{2}>$ (circles)
versus chain length $L$. The fitting functions (linear in L) assume ideal chain statistics 
with prefactors $1.862$ for $<R_g^{2}>$ and $11.06$ for $<R^{2}>$. The fit works for the 
$L>L^{*}$ behaviour. \label{fig5b}}
\end{figure}

Figures~\ref{fig5a} and~\ref{fig5b} show for dilute and semi-dilute state points
respectively, the average of the square of the radius of gyration and the
average of the square of the end-to-end vector as a function of the chain
length within the polydisperse sample.

In figure \ref{fig5a} relative to the dilute case, standard power law scaling $L^{2\nu}$ law 
with $\nu=0.588$ is indicated for long chains. For our polydisperse system with an average
chain length of $L_0 \approx 10$, our data on chains with $L>L_0$ are affected by too 
large statistical error bars to test critically the validity of this asymptotic regime and its
exact range of application, limited in the low L regime by usual short chain effects clearly 
visible below $L=25$.

The conformational properties of chains in the semi-dilute system are compatible with 
ideal chains for $L>L^{*}\approx 12$ where $L^{*}$ is estimated from eq.(\ref{eq:blo}) 
with $\phi=0.15$, and with a larger exponent power law for lower L values but poor 
statistics and small chain effects prevent a firm test of scaling laws for these 
semi-dilute chains.

When $<R^{2}>$ data from both state points are plotted versus $L$ on a unique graph 
(not shown), the short chains behaviours are found to be undistinguishable, while divergences 
between the two state points are observed for long chains. The power laws shown 
in figures~\ref{fig5a} and~\ref{fig5b} intersect at the blob length value of 
$L^{*}=37$ monomers, giving a blob size or a correlation length of 
$\chi=<R^{2}>^{1/2}=20$. Note that in this analysis, we implictly assume that the prefactor
of the dilute chains power law is independent of $\phi$ while we consider the $\phi$
dependent prefactor in the ideal chain power law corresponding to the semi-dilute state 
at $\phi=0.15$. We thus see that the blob size as determined from the 
intersection of asymptotic power laws (neglecting short chains effects) suggest a 
numerical prefactor of about $3$ in the scaling relations (\ref{eq:blo}). On this basis,
our dilute system at $\phi=0.05$ would be characterized by a blob length of $L^{*}=50$ and
thus a ratio $L_0/L^{*}=0.2$ suggesting a rather clear dilute solution character. The same ratio 
is $L_0/L^{*}=1.4$ for the more concentrated solution showing that the usual concept of chains of 
blobs in a semi-dilute solution applies only marginally to the longer chains in the sample. 
To conclude, this analysis indicates that, even in the semi-dilute case, the chain dynamics 
should not be significantly affected by the weak entanglements between chains. Let us stress that 
the semi-dilute character of the sample is however well illustrated by static properties
such as the exponential chain length distribution and the ideal chain power laws for chain sizes.

\section{Kinetics analysis}

The number of Monte-Carlo ``accepted" scissions (or ``accepted" recombinations) 
are obtained by simple counting during the simulations. In table~\ref{table1}
(dilute case) and in table~\ref{table2} (semi-dilute case), we list the number 
$n_s$ of ``accepted" transitions per unit time and unit volume, for the two 
state points and for the different attempt frequencies $\omega$ investigated. 
For each state point, we observe that the number of scissions per unit of 
time is, as expected, proportional to $\omega$ and we have also verified that 
the number of recombinations differs from the number of scissions by marginal 
amounts (0.05$\%$), which shows that the chain length distributions are well
equilibrated.

\begin{table} 
\caption{Kinetic data for the dilute case at four scission-recombination attempt
frequencies $\omega$. $n_s$ is the total number of Monte-Carlo scissions 
(recombinations) per unit time and per unit volume. $\alpha$ is the fraction of scissions 
where newly created free arms recombine with a new partner. 
Superscript $(i)$ refers to the methodology used to get various 
estimates of a given quantity, namely $i=1$ for the method based on the 
distribution of first recombination times, $i=2$ for the cumulative hazard method and 
$i=3$ for the ``reactive flux" approach. $\tau_b$ is the average first recombination 
time and also the typical life time of a chain of average length $L_0$. The estimate 
$\tau_b^{(i)}$ is obtained by the long time behaviour of the relevant dynamical function
used in methodology (i). $\kappa^{(i)}$ is the fraction of transitions which are effective.
Such transitions can be seen as those which do not belong to sequences of correlated transitions 
(chain scission followed by almost immediate recombination ending into no change in chains topology).
An additional estimate of $\tau_b$ (not listed in th etable) may be obtained on the basis of $n_s$ 
and $\kappa^{(i)}$ computed with methodology $i$ using $\tau_b=\frac{\phi}{L_0 \kappa^{(i)} n_s}$}
\centerline{\begin{tabular}{|c|c|c|c|c|c|c|c|}
\hline   $\omega$ & $\kappa^{(3)}$ & $\alpha$ & $n_s*10^{3}$ & $\tau_b^{(1)}$ & $\tau_b^{(2)}$ & $\tau_b^{(3)}$ \\
\hline   0.1 & .89 & 0.89 (0.80) & 0.00759  & 694. & 763. & 625.\\
\hline   0.5 & 0.60 & 0.68 (0.60) & 0.038  & 196. & 212. & 185.\\
\hline   1.0 & 0.45 & 0.55 (0.48) & 0.076 & 153. & 146. & 122.\\
\hline   5.0 & 0.20 & 0.28 (0.25) & 0.38 & 60.2 & 67.0 & 63.3\\
\hline
\end{tabular}}
\label{table1}
\end{table}

\begin{table} 
\caption{Kinetic data for the semi-dilute case at four scission-recombination
attempt frequencies.All quantities are defined in the previous table} 
\centerline{\begin{tabular}{|c|c|c|c|c|c|c|c|}
\hline   $\omega$ & $\kappa^{(2)}$& $\kappa^{(3)}$ & $\alpha$ & $n_s*10^{3}$ & $\tau_b^{(1)}$ & $\tau_b^{(2)}$ & $\tau_b^{(3)}$ \\
\hline   0.1 & 0.85 & 0.85 & 0.85 (0.73) & 0.0042 & 919. & 885. & 813. \\
\hline   0.5 & 0.52 & 0.44 & 0.53 (0.44) & 0.022 & 285. & 265. & 282.  \\
\hline   1.0 & 0.34 & 0.35 & 0.40 (0.31) & 0.042 & 191. & 190.& 184  \\
\hline   5.0 & 0.12 & 0.12 & 0.16 (0.11) & 0.21 & 130. & 108. & 106.  \\
\hline
\end{tabular}}
\label{table2}
\end{table}

The quantity $\alpha$ mentioned in the tables is the fraction of scissions
which lead to a recombination with a new partner. In our analysis, the fate of 
a scission has been initially followed for a maximum time of $T_{max}=1000$. 
The value quoted in parentheses is the ratio obtained by using data for which 
the recombination time is less than $T_{max}$, while the first value is obtained 
by assuming that all recombinations taking place beyond that maximum time should 
be necessarily with new partners, as it is justified by data shown in figure 
\ref{fig6b} to be discussed later. The product $\alpha n_s$ is the number of 
scissions per unit of time and per unit volume which lead to a 
recombination with a new partner, that is those which lead to real changes in 
the distribution of chain lengths.

In the explored $\omega$ range, the value of $\alpha$ indicates that a large 
fraction of elementary scissions are followed by recombinations of the same 
original chain fragments. Using $\alpha$ as a measure of the percentage of 
effective reactions, we observe a decrease of this percentage as the rate 
of attempted transitions $\omega$ increases. Actually, uneffective transitions 
may also come from more complex particular transition sequences. Consider a 
chain end monomer (say monomer $i$) lying close in space to another chain at 
the level of two adjacent monomers $j$ and $k$. At high $\omega$, a scission 
of the bond $jk$ may be immediately followed by a recombination between $j$ or 
$k$ with monomer $i$. In turn, that $ik$ or $ij$ bond may reopen and recombine 
to restaure the very first situation, ending with no effective transition 
without being detected through the criterium of a successive recombination with 
the same partner. The occurence of such events is proven indirectly in the following
by noting almost immediate chain end recombination with another partner (see figure
\ref{fig6b}).

To extract estimates of the rate constants defined by the kinetic model of
Cates, it is thus important to eliminate the spurious transitions from the effective 
ones. The best route for this is certainly to approach the problem via a time scale
separation between scission-recombination processes taking place on a
fast time scale and effective transitions taking place on the reaction time
scale $\tau_b$.

\subsection{Distribution of first recombination times}

\begin{figure}[th] 
\centerline{\includegraphics[width=.75\textwidth,angle=-90]{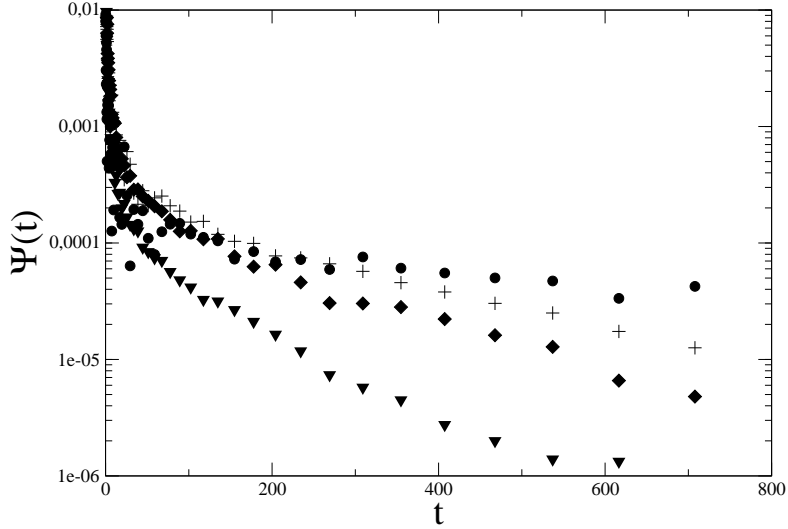}}
\vspace*{8pt}\caption{Distribution of first recombination times $\Psi(t)$ for
the semi-dilute case. Data are shown for $\omega=0.1$ (filled circles), $\omega=0.5$ 
($+$ symbols), $\omega=1.0$ (filled lozenges) and $\omega=5.0$ (filled triangles).
Estimates of $\tau_b^{(1)}$ are obtained from the slope of a linear fit of 
$\ln{\Psi(t)}$ versus time in the long time region.
\label{fig6a}}
\end{figure}

\begin{figure}[th] 
\centerline{\includegraphics[width=.75\textwidth,angle=-90]{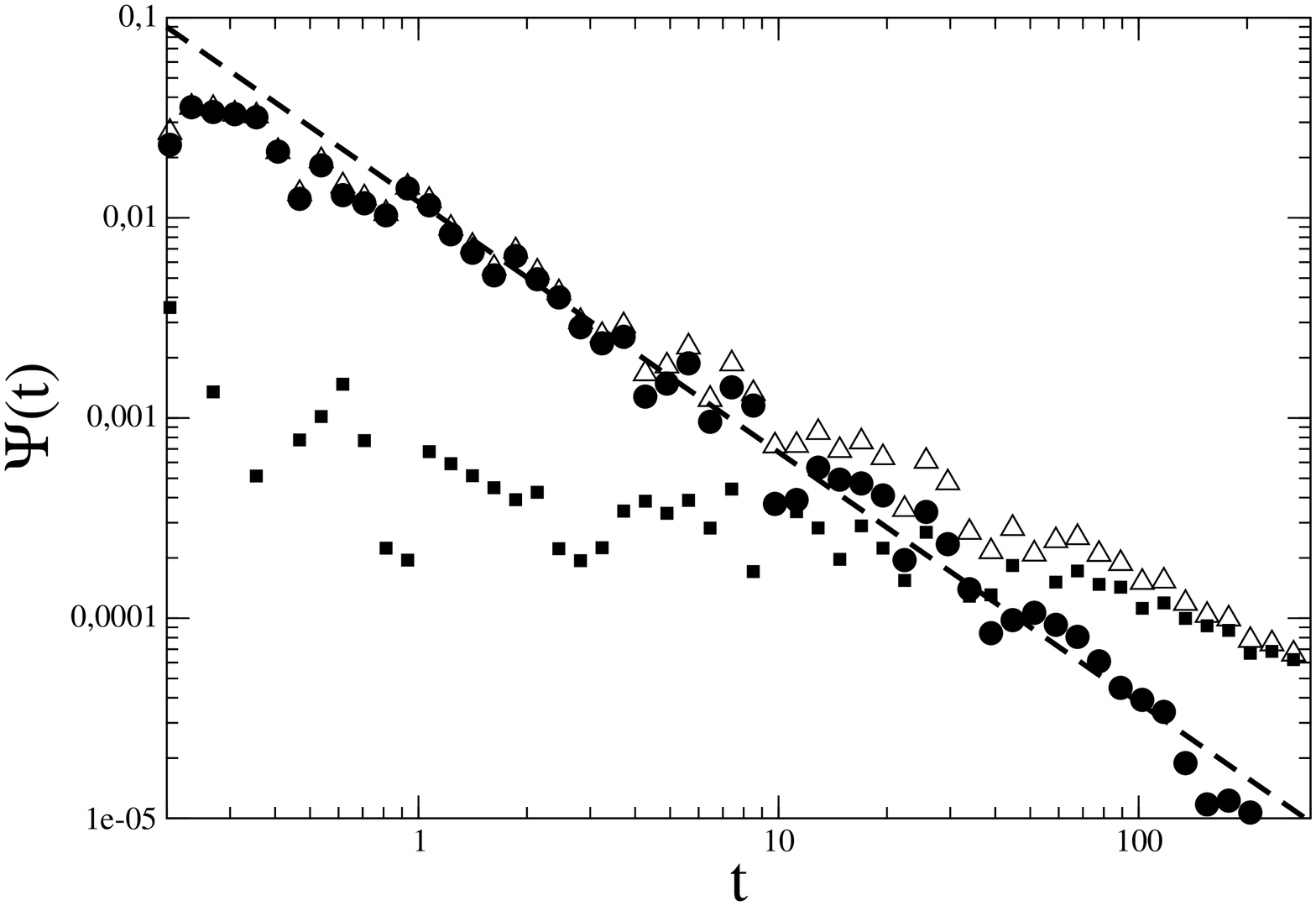}}
\vspace*{8pt}\caption{Distribution of first recombination times $\Psi(t)$ for
the semi-dilute case for attempt frequencies $\omega=0.5$ (open triangles). The partition
between contributions from recombinations with the same partner (filled circles) or with a 
new partner (filled square) are distiguished. The power law $0.012 (t)^{-5/4}$ shows 
that the ``diffusion-controlled" theoretical prediction of $-5/4$ for the power law exponent 
is perfectly satisfied for the self recombination part for times larger than $t=1$. 
The picture shows how self-recombinations dominate at short times while recombinations 
with new partners dominate beyond $t=30$.
\label{fig6b}}
\end{figure}

The first approach is to compute the histogram of first recombination times
$t=t_2-t_1$ for an arm which became free by scission at time $t_1$ and which
recombined with another free arm at time $t_2$. Figure~\ref{fig6a} shows the
result for the semi-dilute case for the four values of $\omega$. In
figure~\ref{fig6b} relative to the case $\omega=0.5$ specifically, we show the
contribution to the distribution of first recombination times from events
implying either the same or a different partner. In the latter figure, one
observes that at short time, the recombination with the same partner 
dominates. However, even at short times, recombinations with another partner are 
possible as discussed earlier. The two curves corresponding to the two types of 
contributions cross each other around $t=25$ and at later times, the recombination 
with another partner progressively dominates. We note however that at times as long as
$t=200$ where monomers have diffused by a distance corresponding to three times
their size (see figure 9), there is still $10\%$ of recombinations at that time
which take place with the same partner.

Data of figure~\ref{fig6a} have been analyzed by considering that $\Psi(t)$ is
exponential at long times. The slope is interpreted as the inverse of the
average life time of a chain end which is denoted as $\tau_b^{(1)}$. Specific
values for the different $\omega$'s are indicated in tables 1 and 2, respectively for dilute and 
semi-dilute cases.

The short time behaviour of the distribution can be analyzed by plotting the function versus
time on a logarithmic scale. This is done in figure~\ref{fig6b} where we distinguish 
contributions from recombinations with the same partner or with a new partner. The theory
of diffusion controlled recombination kinetics\cite{oshau} predicts an algebraic decay
$A t^{-5/4}$  for $\Psi(t)$ and we observe that it is perfectly satisfied by our data
for the self recombination part for times larger than $t=1$. The picture shows how 
self-recombinations dominate at short times while recombinations with new partners 
dominate beyond $t=30$. At very short times (t<1), a significative part (about $10\%$) of
the $\Psi(t)$ function implies recombinations with anorther partner.

\subsection{Cumulative hazard analysis}

To analyse the rates of conformational transitions in butane and other short
alkane molecules, Helfand \cite{helf} suggested to exploit hazard rate
plotting. We have adapted this technique to the present case. We first summarize the 
theoretical foundations of the method, using the particular case of equilibrium 
polymer kinetics to directly illustrate the concepts.

Let $h(t)dt$ be the probability that a free arm, created at time $t=0$ and which is still 
free at time t, undergoes a first recombination in the interval $[t,t+dt]$. Let $P(t)$ be the 
probability that a free arm, created at time $t=0$, has undergone a (first) recombination
between 0 and t.

Using these definitions, the following steps can be written
\begin{eqnarray}
\left[1-P(t+dt)\right]&=&\left[1-P(t)\right](1-h(t)dt)\\
\frac{d}{dt}\left[1-P(t)\right]&=&-h(t)\left[1-P(t)\right]\\
\left[1-P(t)\right]&=&\exp{(-\int_0^{t} h(t')dt')}=\exp{\left[-H(t)\right]}
\end{eqnarray}
where $H(t)=\int_0^{t} h(t')dt'$ is the cumulative hazard.  In the present case, 
we anticipate a complex process involving correlated events at short times 
$h(t)=h^{sh}(t)$ and a Poisson process emerging at long times with uniform frequency 
$lim_{t\rightarrow\infty}h(t)=\lambda$ where $\lambda$ is the recombination rate constant.
On the relevant time scale of the kinetics, one thus should find for the cumulative hazard 
function and the P(t) probability
\begin{eqnarray}
H(t)&=&H_I+\lambda t\\
P(t)&=&1-\exp{(-H_I)} \exp{(-\lambda t)}
\label{eq:poisson}
\end{eqnarray}
where $H_I$, the ordinate intercept of the function $H(t)$ versus t, can be seen as the time integral 
of $(h^{sh}(t)-\lambda)$ from $0$ to $\infty$. If a good time scale separation exists, eq.(\ref{eq:poisson}) implies that $c \equiv P(0^{+})=1-\exp{(-H_I)}$ is the fraction of correlated transitions, $\kappa^{(2)}=1-c$ being the estimate according to the present analysis of the average probability for a newly created free arm to 
recombine by the Poisson ``mean field" kinetic process postulated by Cates in his theory\cite{cates}.

We now explain how the estimate of the cumulative hazard function is constructed. We start by 
extracting from the BD trajectory a collection of times  $\{\hat{t}\}$ where each member 
corresponds to the elapsed time between a scission of a particular arm at time $t_1$ and 
its next recombination at time $t_2$, so $\hat{t}=t_2-t_1$. All the $2M$ arms of the 
system contribute to the $\hat{t}$ data sample which, for each arm, can contain several times of 
this kind between the start (at $t=0$) and the end (at $t=T_{max}$) of the BD trajectory.

Moreover, if the first change of status of a particular arm since the beginning
of the BD simulation is a recombination taking place at time t, we can say that
this time is a lower bound $\hat{T}$ of an additional unknown elapsed time
$\hat{t}$ between a scission (out of our reach) and the next recombination (we
observed). Also, if the last change of arm status before the end of the BD
trajectory is a scission taking place at time t, then, the time
$\hat{T}=T_{max}-t$ is a similar lower bound of yet another time of
interest. The analysis thus furnishes a set of $K$ times $\{\hat{t}\}$ and $M$
lower bounds times $\{\hat{T}\})$ which are then separately ordered from the
shortest time up to the longer one and indexed accordingly as $(\hat{t}_1,
\hat{t}_2, \hat{t}_3,..\hat{t}_{K})$ and $(\hat{T}_1, \hat{T}_2,
\hat{T}_3,...\hat{T}_{M})$. Let us consider successively all individual arm
life times $\{\hat{t}\}$. For any time $\hat{t}_i$ of that collection, the
probability that a free arm which had survived up to the previous time
$\hat{t}_{i-1}$ changes its status and becomes engaged in the formation of new
bond between times $\hat{t}_{i-1}$ and $\hat{t}_i$ is given with our available
statistics by $1/N(\hat{t}_{i-1})$ where the denominator is the number of cases
(including both types of times $\hat{t}$ and $\hat{T}$) where recombination
takes place at a time longer than $\hat{t}_{i-1}$, i.e.

\begin{equation}
N(\hat{t}_{i-1})=[K-(i-1)] + [M-m_{(i-1)}]
\end{equation}

where $m_{(i-1)}$ is the index of largest $\hat{T}$ value which is still
inferior to $\hat{t}_{i-1}$. In terms of these definitions, the cumulative
hazard function $H(t)$ can be evaluated at each time $\hat{t}$

\begin{equation}
H(\hat{t}_i)=\sum_{j=1}^{i-1} \frac{1}{N(t_j)} \label{eq:cumul}
\end{equation}

If the function is linear in time, it implies a Poisson process with a rate of
transitions given by the slope of that linear portion.

\begin{figure}[th] 
\centerline{\includegraphics[width=.75\textwidth,angle=-90]{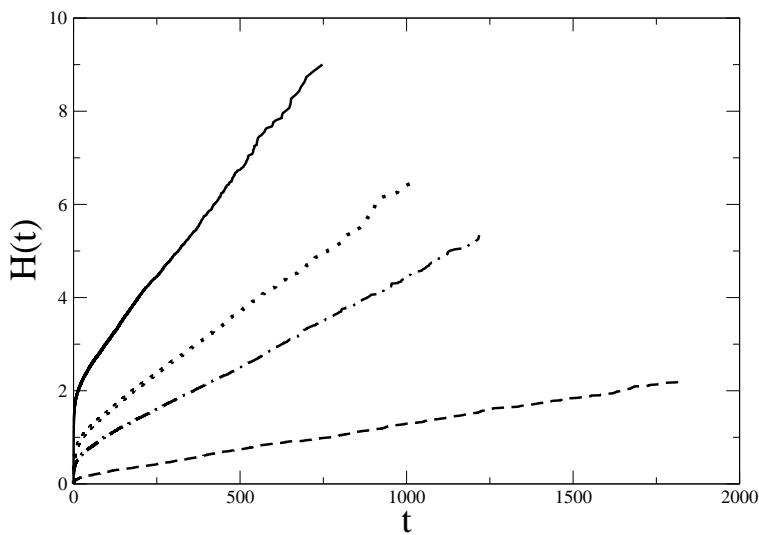}}
\vspace*{8pt}\caption{Cumulative Hazard curves of recombination times for the
semi-dilute case. Data are shown for various attempt frequencies, namely $\omega=0.1$ 
(dashed line), $\omega=0.5$ (dot-dashed line), $\omega=1.0$ (dotted line) and
 $\omega=5.0$ (continuous line).\label{fig7}}
\end{figure}

Figure~\ref{fig7} shows the resulting cumulative hazard functions for the four
$\omega$ values in the semi-dilute case. A linear behaviour starts around
$100-150$ in all cases and it lasts up to times of the order of $1000$ where
statistical noise sets in. The inverse slopes of the linear portions provide
estimates $\tau_b ^{(2)}$ of the mean life time of a chain end. Specific values
for both thermodynamic states are gathered in table 1 and table 2. 
They are compatible with the values extracted from the histogram of first recombination
times. Using the procedure described at the beginning of this section, one gets from the 
ordinate intercepts of $H(t)$ the $\kappa^{(2)}$ estimate indicated in table 2 for the 
semi dilute case.

The cumulative hazard plot has the advantage that it avoids the necessity of
bining recombination time data which become scarse at long times. It also exploits
statistically some additional information (useful for low rates) from portions
of the $2M$ trajectories between a scission and the next recombination in a way
which is truncated either at the beginning or at the end of the BD run.

\subsection{The reactive flux correlation function and the transmission coefficient}

We now discuss still another way to isolate short time transitional effects
in order to estimate the rate of ``effective" transtions. It relies on the way
so-called ``Transition State Theory" (TST) values of chemical reactions rates
can be corrected by a multiplicative transmission coefficient $0<\kappa<1$
which takes into account the fraction of events where the products formed by
the forward reaction are effective transitions. By effective, it means that
after such a transition, the product is able to relax its energy excess (which
was accumulated to cross the forward barrier) sufficiently quickly to avoid an
immediate recrossing of the barrier in the backward direction.

In our case, the analogy is clear. If a bond scission is followed by the same
bond recombination within a short microscopic time scale (too short for
relative diffusion to take place), it basically means that no effective
scission has happened to the chain to which the bond belongs on time scales
larger than the relaxation time of a chain fragment of a few monomers.

We have thus estimated $\kappa^{(3)}$ by exploiting the concept of the reactive flux
correlation function, as explained by Chandler in the framework of an
equilibrium between forward and backward unimolecular reactions\cite{chand}.
We assume here that for an equilibrium between forward unimolecular reactions
(chain breaking) and bimolecular backward reactions (end-chain recombinations),
the same conceptual expression can still be applied for the forward reaction.

Let us consider one specific arm (among the two of one particular monomer)
which is labelled by an index $i$ and to each arm, we associate a signature
variable $S_i(t)$ which is equal to unity if the ith arm is free at time $t$
and equal to zero if the arm is engaged in a binding with another monomer at
that time. Crossings are then characterized by a step change of the $S_i(t)$
values, $S_i(t)$ going from 0 to 1 or from 1 to 0 for scissions and
recombinations respectively. A statistics over all crossing events is defined:
for each individual crossing, we set the time at crossing to zero, denoting
times very shortly after or before the transition by $t=0^{+}$ and $t=0^{-}$.
The reactive flux correlation function is then defined by\cite{brown}

\begin{equation}
C_{RF}(t)= 2 <\left[S_i(0^{+})-S_i(0^{-})\right] S_i(t)>_{cross}
\end{equation}

where the average is performed over all crossings (both scissions and
recombinations). If we treat separately the averages on scissions and
recombinations, we get the equivalent alternative formulation

\begin{equation}
C_{RF}(t)=  < S_i(t)>_{scissions}- < S_i(t)>_{recombinations}
\end{equation}

This function starts from unity at very short (positive) times with the first term
being one and the second being zero. On very long times (long respect to the
relevant inverse kinetic rate constants), this function goes to zero as it is
the difference of two averages which ultimately converge to the same
equilibrium average value $<S_i>$.

\begin{figure}[th] 
\centerline{\includegraphics[width=.75\textwidth,angle=-90]{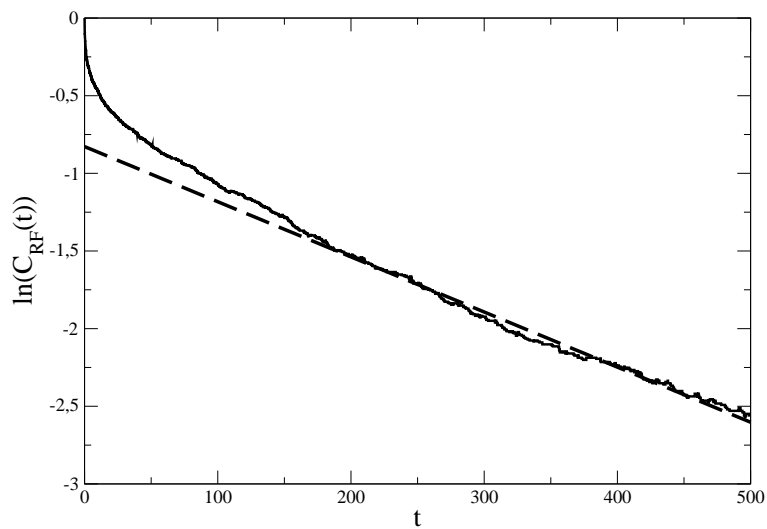}} \vspace*{8pt}\caption{
Reactive flux correlation function in the semi-dilute case for $\omega=0.5$. The dashed
line corresponds to a fit by a function $\kappa \exp{(-t/\tau_b)}$ in the time region 
$150<t<500$. Best parameters found are $\kappa^{(3)}=0.48$ and $\tau_b^{(3)}=282$.
\label{fig8}}
\end{figure}

If all transitions are effective, the reactive flux function will decrease
smoothly from 1 to zero reflecting only the kinetics of these effective
recombination and scissions. If transitions are often followed by opposite
transitions within a short time scale $\tau_{micro}$ or if successive
transitions are correlated within such a short time, this will be reflected in
the reactive flux by a fast initial decay over that microscopic time. At later
times, the macrocopic kinetics (the one considerd by mean field theoretical
expresions) will be reflected by the $C_{RF}(t)$ evolution towards zero on a
macroscopic time scale, $\tau_{macro}$. The transmission coefficient $\kappa^{(3)}$
is thus estimated as the initial value of the ``slow" decay part of that
$C_{RF}(t)$ function. This can be expressed as
\begin{equation}
\kappa^{(3)} = C_{RF}(\tau_{micro})
\end{equation}
with the requirement of a good time scale separation, that is $\tau_{micro} \ll
\tau_{macro}$.

Figure~\ref{fig8} shows the $\omega=0.5$ case for the semi-dilute case already
discussed in fig~\ref{fig6b} in the context of the distribution of first
recombination after a scission. This curve is more difficult to analyse because
it is difficult on the basis of this sole function to decide when the behaviour
of $ln(C_{RF}(t))$ really becomes close from a linear behaviour. We had to rely
on our experience with the distribution $\Psi(t)$ and the cumulative hazard
function $H(t)$ for that state point and $\omega$ value. As both functions
suggest that the simple ``mean field" process emerges at times later than 150,
we have fitted the $C_{RF}(t)$ data by an exponential function
$C_{RF}^{fit}(t)=\kappa^{(3)}*\exp{(-t/\tau_b^{(3)}})$ in the accessible time window
above $t=150$. In the present case, we see that there is no real time scale
separation between a short time behaviour, observed in the \linebreak$0<t<150$ regime,
and the macroscopic time scale $\tau_b^{(3)}$ which is equal to $\approx 280$.
Nevertheless, the value of $\tau_b^{(3)}$ turns out to be rather close to the
other estimates of a chain end average life time and the $\kappa^{(3)}$ value will be 
shown in the next subsection to lead to a reasonable correction of $n_s$, the total 
number of transitions per unit of time and unit volume quoted in tables 1 and 2. 
Results of $\kappa^{(3)}$ and $\tau_b^{(3)}$ for all $\omega$'s and both state points 
are included in the tables 1 and 2.

\subsection{Estimation of rate constants: comparison of the various methods}

Explicitly, the scission rate constants $k_s$ and $k_r$ of the ``mean-field" kinetic model can be
estimated from our data in two distinct ways.

\begin{itemize}
\item On the basis of the total number of transitions $n_s$ per unit time and per unit
volume, and on the basis of an estimate of $\kappa$ by one of the methods discussed above,
the rates can be estimated from a trivial modification of eqs.(\ref{eq:ntran1},\ref{eq:ntran2}), 
namely
\begin{equation}
\kappa n_s= k_s \phi = \frac{k_r \phi^{2}}{2 L_0^{2}}
\label{eq:ntranc}
\end{equation}
\item On the basis of the long time behaviour of a dynamical function relevant to the 
particular methodology adopted, the ``chain end" life time can be directly estimated. As 
the latter is equivalent to the typical life time of a chain of average length denoted as
$\tau_b$, one gets rate constants through the equivalence
\begin{equation}
\tau_b=\frac{1}{k_s L_0}=\frac{2 L_0}{k_r \phi}
\label{eq:lifetime}
\end{equation}
where the last equality follows from detailed balance requirements. 
\end{itemize}

Tables 1 and 2 provide all needed data. We find that the implicit time scale separation 
inherent to all analyses provide compatible estimates of the effective rate constants. This can
be verified by applying eqs (\ref{eq:ntranc}) and (\ref{eq:lifetime}) with the particular 
estimate of the left hand side.

Looking at all values in the tables, we get an overall consistency between all methodologies.
It indicates that all our strategies to extract the ``macroscopic" rate constants do work. 
Among them, the cumulative hazard analysis appears the most straightforward to analyse once 
$H(t)$ is known and particularly robust as noise influence seems minimal. 

We are now in a position to compare the percentage of effective transitions $\kappa$
based on a time scale separation and the $\alpha$ value based on the percentage of
recombinations involving a new partner respect to the one left by scission. We
observe in the two tables that both estimates are reasonably consistent and quickly 
decrease for $\omega$ increasing. Fine analysis suggests however that some distinction 
exists between both quantities in the present case, that is a free Rouse relaxation type of 
dynamics in solution for moderately long chains. It requires further analysis and the consideration 
of more entangled cases to make progress on that point.
 Finally, it should be stressed that while diffusion controlled and mean field kinetics have 
 sometimes been opposed to each other\cite{padd}, it appears in our analysis that they consistently describe 
 short time correlation effects dominated by self recombinations and long time kinetics dominated by 
 recombinations with new partners.

\subsection{Analysis of the monomer diffusion}

\begin{figure}[th] 
\centerline{\includegraphics[width=.75\textwidth,angle=-90]{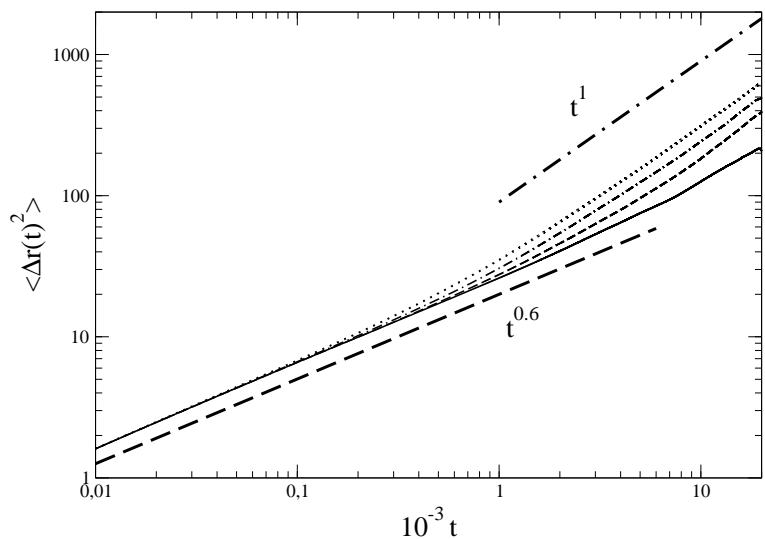}}
\vspace*{8pt}\caption{Mean square displacement of monomers (semi-dilute case). The 
continuous curve has been obtained for the relevant polydisperse sample but with $\omega$ 
set to zero, averaging over many independent initial configurations. It is a reference 
from which the other curves depart at decreasing times when increasing values of $\omega$ are 
considered. The cases $\omega=0.1$ (dashed lines), $\omega=0.5$ (dot-dashed lines) and
$\omega=5.0$ (dotted lines) are shown together with two indicative power laws which suggest the
evolution of the subdiffusive scaling regime to the normal diffusion limit.
\label{fig9_1}}
\end{figure}

\begin{figure}[th] 
\centerline{\includegraphics[width=.75\textwidth,angle=-90]{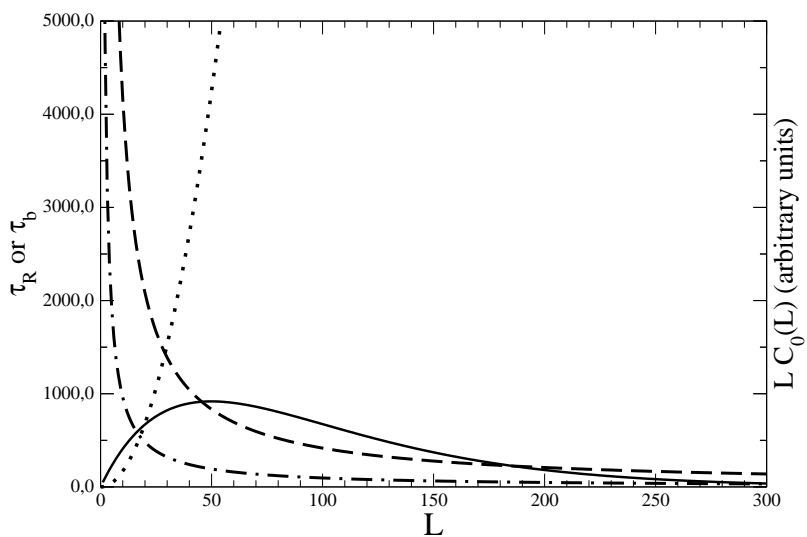}}
\vspace*{8pt}\caption{Characteristic times as a function of polymer size. The Rouse 
relaxation time $\tau_{R}=1.70*L^{2}$, shown as dotted curve, must be compared to the life time
of a polymer of size $L$, as given $\tau_b=(k_s L)^{-1}$. Two frequencies are illustrated, namely 
$\omega=1.0$ (dot-dashed line) and $\omega=0.1$ (dashed line) using effective $k_s$ values. 
We show on the same graph the distribution of monomers (continuous curve) as a function 
of the size of the polymer they belong to (see text).
\label{fig9_2}}
\end{figure}

The mean squared displacement (MSD) of all monomers as a function of time is
shown in Figure \ref{fig9_1} for all $\omega$'s in the semi-dilute case. As
these monomers belong to a polydisperse set of chains which continuously break
and recombine, the interpretation of the MSD is rather complex.
\begin{itemize}
\item Being a single monomer property, a relevant quantity is the average
fraction of monomers belonging to chains of length L, namely a distribution
$\propto L*c_0(L)$. Fig~\ref{fig9_2} shows the resulting curve for the adopted
semi-dilute conditions with $<L>=49.9$.
\item The time scale for the scission/recombination process has been estimated so
far in terms of the average survival time $1/(k_s <L>)$ for the polymer of
average size, a quantity which was found to vary from $\approx 100$ up to
$\approx 900$ in the investigated $\omega$ range (see table 2). In the present
context, we need to consider the explicit L dependent average survival time
$\propto 1/(k_s L)$ for a given $\omega$ value and figure~\ref{fig9_2} shows the
resulting functions for two such frequencies.
\item To discuss the monomer mean squared displacement, it is also important to
have an estimate of the longest internal relaxation (Rouse) time of the chains.
Performing a separate brownian dynamics run on a monodisperse sample of 20
``dead" polymers of length L=50 (corresponding to the mean polymer length of our
living polymer sample) at the same semi-dilute state point ($T=1$,
$\phi=0.15$), we get from the long time exponential bevaviour of the end-to-end vector
relaxation function a value of $\tau_{R}=4.24\ 10^{3}$. For our very flexible
polymers, the scaling of the Rouse time with L in the semi-dilute conditions
should scale as
\begin{equation}
\tau_{R}(L)=A L^{2}.
\end{equation}

On the basis of our estimated value for $L=50$ chains, we get $A \approx 1.70$,
a value which can be compared (at least its order of magnitude) to theoretical estimate
\cite{gennes,roumil} $A \approx \tau_{blob}/(L^{*})^{2}$ in terms of
$\tau_{blob}$, the blob relaxation time, and $L^{*}$, the number of monomers
per blob. We get $A=5.8$ for our conditions (see discussion in the section 
on chain conformational properties) where the number of monomers per blob 
$L^{*}\approx 37$ and $\tau_{blob} \approx \frac{\eta_s \chi^3}{T} \approx 8000$ 
using a blob size $\chi=20$ and a solvent viscosity $\eta_s=\frac{\xi}{3 \pi
\sigma}=1$ estimated from the friction coefficient using the Stokes law with 
an hydrodynamic radius fixed to $\sigma/2$. 
\end{itemize}

On the basis of the above considerations, it is useful to define a particular
chain length $\Lambda$ for which the Rouse time is equal to the survival time,
giving $\Lambda=(1.7*k_s)^{-1/3}$, that is $\Lambda=30, 20, 18, 15$ for
$\omega=0.1,0.5,1.0, 5.0$ respectively. Chains longer than $\Lambda$ have their
internal dynamics strongly altered by scissions, while chains shorter than
$\Lambda$ should have an internal dynamics little affected with respect to dead
polymers of the same size. As figure~\ref{fig9_2} illustrates, when the
scission attempt frequency increases, the fraction of monomers which belong to
chains whose dynamics is little affected by scissions decreases.

\begin{figure}[th] 
\centerline{\includegraphics[width=.75\textwidth,angle=-90]{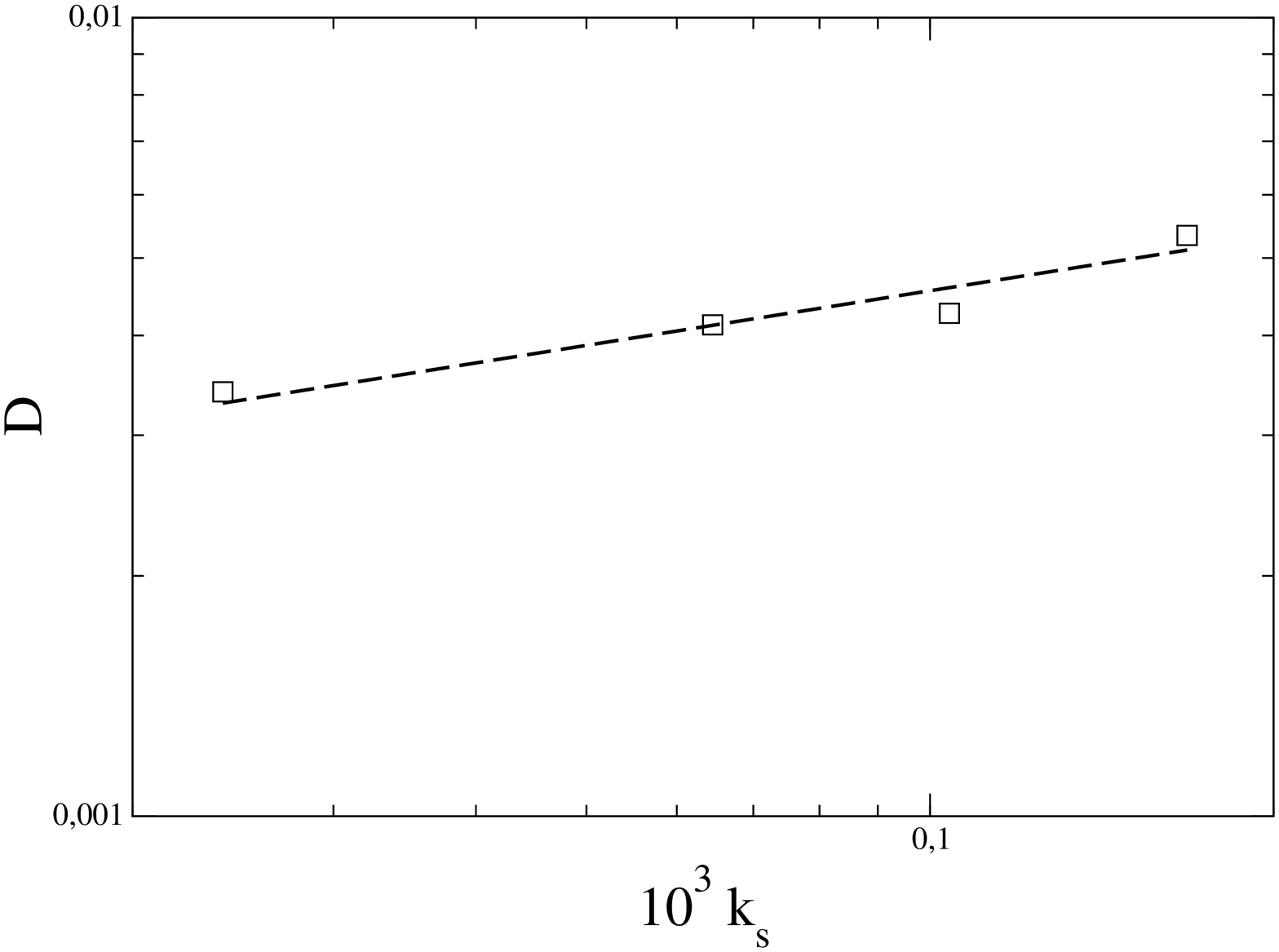}}
\vspace*{8pt}\caption{Monomeric Einstein diffusion coefficient as a function of 
$k_s$ for the semi-dilute case. The value of $D$ was obtained from the long time behaviour (beyond t=8000) of the monomeric mean square displacement shown in figure \ref{fig9_1}. The dashed line represent a $k_s^{0.23}$ power law. \label{fig9_3}}
\end{figure}

Up to a time of 100, the MSD shown in figure~\ref{fig9_1} is $\omega$
independent and presents a Rouse like behaviour with a time scaling exponent
$t°^{0.6}$. In the reference case where $\omega$ was set to zero (we simulate
polydisperse sample of dead polymers), this power law bahaviour persists at
least up to time 10000, a time corresponding to the Rouse time of a chain of
length $170$. When scissions and recombinations are allowed, the MDS deviates
from the master curve (of dead polymers) to adopt progressively a MSD linear in
time. This takes place sooner and sooner for increasing values of $\omega$. At
long times ($t>2000$), the MSD has evolved towards usual Einstein diffusion
with a diffusion coefficient increasing with $k_s$. In figure~\ref{fig9_3}, D is
observed to (very roughly) scale like $k_s^{0.23}$ in the explored range, a
result which is not so far from the $1/3$ power behaviour observed by Milchev
with the lattice BFM lattice \cite{milch}, but which requires further analysis
with better estimates of $D$ as statistics is poor above t=10000. The $1/3$ power
law scaling is explained by assuming that clusters of $\Lambda$ monomers are
responsible of the long time behaviour of living polymers, giving $D \propto
\frac{R_g^{2}(\Lambda)}{\tau_{Rouse}
(\Lambda)} \propto \Lambda^{-1} \propto
k_s^{1/3}$.

\subsection{Analysis of the first recombination times distribution and
corresponding diffusive steps}

\begin{figure}[th] 
\centerline{\includegraphics[width=.75\textwidth,angle=-90]{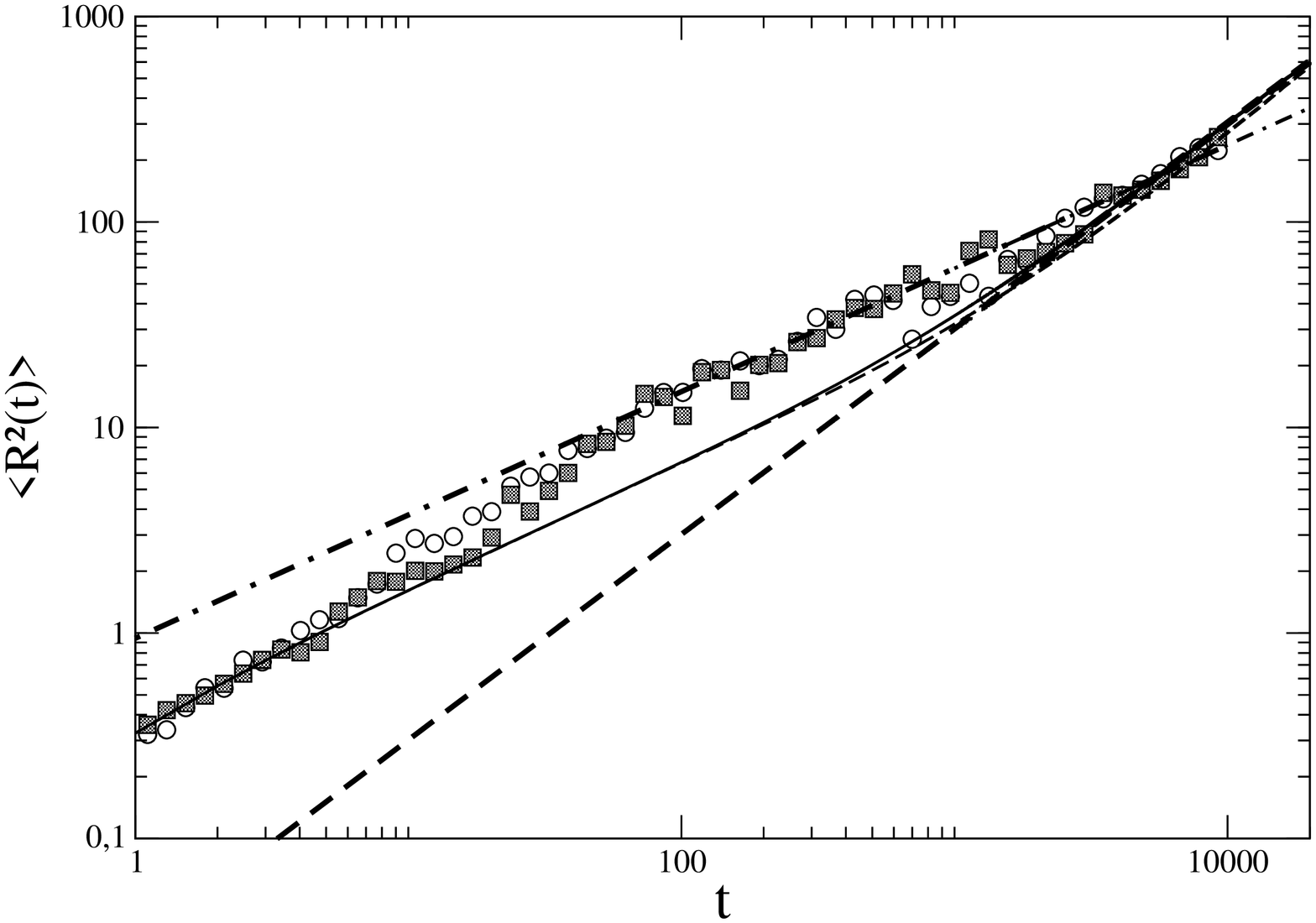}}
\vspace*{8pt}\caption{Mean square displacement of end chain monomers between
a time set to zero when produced by scission up to a time $t$ corresponding to their 
first recombination, as a function of their total life time $t$. Due to 
poor statistics,only data for $\omega=1$ (filled squares) and $\omega=5$ (empty circles)
are shown and are limited to $t=10000$ maximum. Data on the mean square displacement of all monomers for corresponding $\omega$ 
values already shown in figure \ref{fig9_1} are shown again here for comparison (see text) up to a maximum time of 20000. Scaling power laws $1.0 t^{1.0}$ (thick dashed line) and $0.943 t^{0.6}$ (thick dot-dashed line) shown up to t=20000 are used in the text.  \label{fig10}}
\end{figure}

In figure~\ref{fig10}, we report for two different bond change attempt
frequencies $\omega$, the mean squared displacement of a chain end between its
creation and its next recombination, as a function of the corresponding life time
of that chain end. Interestingly, we observe a common behaviour of this
dependence for both frequencies, meaning that the diffusive dynamics of chain
ends is little sensitive to the scission/recombination processes of the chains
as a whole.
 
As chain ends have very different life times for different $\omega$'s, the
typical average distance travelled by a chain end between its creation and its
recombination is also function of that frequency. Mean travel distances of 3.8, 4.6
and 6.8 are obtained for the average chain end life times at $\omega=$5.0, 1.0
and 0.1 respectively, to be compared with the mean static distance between
chain ends (assuming no spatial correlations) given by $h=(<L>/2
\phi)^{\frac{1}{3}}$ which turns to be $h=5.5$ in our semi-dilute case. By
comparison, the radius of gyration of the average chain length in the system is
$R_g(<L>)\approx 10$. The dynamic picture is thus that chain ends diffuse
similarly with time in an environment of other chains which continuously break
and recombine. As the reacting frequency increases, the probability of a chain
end to recombine with a neighbouring monomer increases and thus takes place
earlier, after diffusion over a shorter distance.

\section{Summary and conclusions}

In this paper, we have proposed a new version of a mesoscopic model of equilibrium 
polymers in solution and we have exploited this model to study the structure and 
the dynamics of these complex systems at a dilute and at a semi-dilute state 
point.

The dynamics at the semi-dilute state point has been more specifically
investigated.  Our system consists of chains with an average number of $L_0
\approx 50$ monomers which can be viewed as a collection of blobs of $\approx
37$ monomers. Given the large flexibility of the model and the moderately 
semi-dilute character, our polydisperse polymer solution should be in a 
non-entangled dynamic regime, with a chain relaxation time growing with chain 
length L as $L^2$.

Working with a unique chain length distribution governed by the structural and
thermodynamic microscopic parameters, we have analyzed the specific influence
of the Monte Carlo binding/unbinding change attempt frequency on the dynamical
properties. Mean field theories are found to be valid for times of the order
and beyond the mean life time of a chain of average size provided the kinetic
constants computed from the Monte Carlo accepted binding/unbinding changes are
rescaled by a transmission coefficient which is a measure of the fraction of
successful scissions. This transmission coefficient, when estimated on the basis
of a time scale separation, turns out to be close to the fraction of
recombinations involving the reunion of a chain end with a new partner and no
longer with the monomer it was attached to previously.

The rate constants have been estimated by different techniques which give
coherent estimates. The cumulative hazard function introduced by Helfand to
compute isomerisation rates in chain molecules has been found to be
particularly efficient and simple to implement. In paticular, we find agreement
between 1) kinetic constant estimates based on the the total number of transitions 
rescaled by the transmission coefficient coming from a diffusion controlled mechanism
and 2) estimates based directly on the time scale of the long time exponential decay 
of relaxation functions (mean field Poisson process). It would be interesting to perform 
so-called T-jump experiments\cite{roumil} with our system to test whether the time 
evolution of the average polymer length $L_0$ towards its new equilibrium value at a different
temperature confirms the mean field analytical prediction with the kinetic constants 
computed in the present work from various population relaxation functions at equilibrium.

The chain size that we chose originally was adequate to demonstrate the validity of the 
methodological aspects developped in this work. In order to test dynamical scaling with 
a wider range of the scission-recombination process frequency, longer chains will be needed
which may require better optimized methods to take solvent effects into account.
The extension of this study towards the rheological properties of our model
system are under way.

\section*{Acknowledgments}
\addcontentsline{toc}{section}{Acknowledgements}

This work has taken profit of numerous discussions with M. Baus of the
Univerist\'{e} Libre de Bruxelles. C.C.H. was financially supported by the Action
de Recherche Concert\'{e}e $N^{0}$ 00/05-257 of the Communaut\'{e} Fran\c{c}aise de Belgique. The 
simulations have been performed thanks to ressources from the Centre de Calcul of the ULB/VUB and the 
French Centre CINES, which we acknowledge.

\renewcommand{\theequation}{A.\arabic{equation}}


\begin{thebibliography}{000}
\bibitem{VdS}
P. van der Schoot, {\sl in ``Supramolecular polymers"} {\it Ed. A. Ciferri}, 77
(N.Y. 2005)
\bibitem{catc}
M.E. Cates and S.J. Candau, {\sl J. Phys: Cond. Matt.} {\bf 2}, 6869(1990)
\bibitem{leroug}
S. Lerouge, J.P. Decruppe and C. Humbert {\sl Phys. Rev. Lett.} {\bf 81},
5457(1998)
\bibitem{hoff}
H. Hoffmann, S. Hoffmann, A. Rauscher and J. Kalus, {\sl Prog. Colloid Polym.
Sci.} {\bf 84}, 24(1991)
\bibitem{OBL90}
A. Ott, J.P. Bouchaud, D. Langevin, W. Urbach,
Phys. Rev. Lett., {\bf 65}, 2201 (1990).
\bibitem{gennes}
P.G. de Gennes, {\sl Scaling concepts in polymer physics} Cornell University
Press, Ithaca(1979)
\bibitem{GK94}
A.Y. Grosberg and A.R. Khokhlov, {\sl Statistical Physics of Macromolecules}
AIP Press, New-York(1994)
\bibitem{doi}
M. Doi and S.F. Edwards, {\sl The theory of polymer dynamics} Clarendon,
Oxford(1986)
\bibitem{cates}
M.E. Cates, {\sl Macromolecules} {\bf 20}, 2289(1987)
\bibitem{spen} N.A. Spenley, M.E. Cates and T.C.B. MacLeish {\sl Phys. Rev. Lett.} {\bf 71},
939(1993)
\bibitem{olms}
S.M. Fielding and P.D. Olmsted {\sl Phys. Rev. Lett.} {\bf 90}, 224501(2003)
\bibitem{oshau}
B. O'Shaughnessy and J. Yu, {\sl Phys. Rev. Lett.} {\bf 74}, 4329(1995)
\bibitem{witt98} J.P. Wittmer, A. Milchev and M.E. Cates {\sl Europhys. Lett.} {\bf 41}, 291(1998);
{\sl J. Chem. Phys.} {\bf 109}, 834(1998)
\bibitem{milch}
A. Milchev, {\sl in ``Computational methods in colloid and interface science"}
{\it Ed. M. Borowko and M. Dekker}, 510(N.Y. 2000)
\bibitem{rouau}
Y. Rouault {\sl J. Chem. Phys.} {\bf 111}, 9859(1999)
\bibitem{padd}
J.T. Padding and E.S. Boek, {\sl Europhys. Lett.} {\bf 66}, 756(2004)
\bibitem{krog}
M. Kroger and R. Makhloufi {\sl Phys. Rev. E} {\bf 53}, 2531(1996)
\bibitem{huang}
C.C. Huang, H. Xu and J.P. Ryckaert, manuscript in preparation,(2005)
%
\bibitem{chand}
D. Chandler, {\sl Introduction to Modern Statistical Mechanics} ,
Oxford University Press (1987)
%
\bibitem{krem}
K. Kremer and G.S. Grest, {\sl J. Chem. Phys.} {\bf 92}, 5057(1990)
\bibitem{MWL99}
A. Milchev, J.P. Wittmer, and D.P. Landau
Monte Carlo Study of Equilibrium Polymers in a Shear Flow
EPJ B, 12, 2, pp. 241-251 (1999); cond-mat/9905336.
\bibitem{MW01}
A. Milchev, J.P. Wittmer, P. van der Schoot, D. Landau
Europhysics Letters, 54 58-64 (2001); conf-mat/0008276.
\bibitem{WM00}
J.P. Wittmer, M.Milchev, P. van der Schoot, J.-L. Barrat
J. Chem. Phys., 113 (2000) 6992-7005; cond-mat/0006465.
\bibitem{helf} E. Helfand, {\sl J. Chem. Phys.} {\bf 69},
1010(1978)
\bibitem{brown}
D. Brown and J.H.R. Clarke, {\sl J. Chem. Phys.} {\bf 93}, 4117(1990)
\bibitem{roumil}
Y. Rouault and A. Milchev, {\sl Eur. Phys. Lett.} {\bf 33}, 341(1996)
\end{thebibliography}
\enddocument